\documentclass[preprint2,times,tighten]{aastex61}
\usepackage{graphicx}

\DeclareGraphicsExtensions{.png,.pdf,.jpg,.ps,.eps}

\newcommand{\mstar}{\ensuremath{{M}_{\star}}}
\newcommand{\rstar}{\ensuremath{{R}_{\star}}}
\newcommand{\teff}{\ensuremath{T_{\rm eff}}}
\newcommand{\fbol}{\ensuremath{F_{\rm bol}}}

\newcommand{\loggstar}{\ensuremath{\log{g_\star}}}
\newcommand{\msun}{\ensuremath{{M}_\Sun}}

\newcommand{\logg}{\ensuremath{\log g}}
\newcommand{\rhostar}{\ensuremath{\rho_\star}}

\newcommand{\rplanet}{\ensuremath{R_p}}
\newcommand{\mplanet}{\ensuremath{M_p}}

\newcommand{\sini}{\ensuremath{{\rm sin}~i}}

\newcommand{\depth}{\ensuremath{\delta_{\rm tr}}}
\newcommand{\ar}{\ensuremath{a/R_\star}}
\newcommand{\rvamp}{\ensuremath{K_{\rm RV}}}

\newcommand{\rprstar}{\ensuremath{R_p / R_\star}}

\newcommand{\flick}{\ensuremath{F_8}}

\newcommand{\nsample}{{675}}
\newcommand{\nsamplefin}{{525}}

\begin{document}

\title{Empirical, Accurate Masses and Radii of Single Stars with {\it TESS\/} and {\it Gaia}}
\author{Keivan G.\ Stassun}
\affiliation{Vanderbilt University, Department of Physics \& Astronomy, 6301 Stevenson Center Ln., Nashville, TN  37235, USA}
\affiliation{Fisk University, Department of Physics, 1000 17th Ave.\ N., Nashville, TN 37208, USA}
\author{Enrico Corsaro}
\affiliation{INAF--Osservatorio Astrofisico di Catania, via S.\ Sofia 78, 95123 Catania, Italy}
\author{Joshua A.\ Pepper}
\affiliation{Lehigh University, Department of Physics, 16 Memorial Drive East, Bethlehem, PA 18015, USA}
\author{B.\ Scott Gaudi}
\affiliation{The Ohio State University, Department of Astronomy, Columbus, OH 43210, USA}

\begin{abstract}
We present a methodology for the determination of empirical masses of single stars through the combination of three direct observables with {\it Gaia\/} and {\it TESS}: (i) the surface gravity via granulation-driven variations in the {\it TESS\/} light curve, (ii) the bolometric flux at Earth via the broadband spectral energy distribution, and (iii) the distance via the {\it Gaia\/} parallax. We demonstrate the method using \nsamplefin\ {\it Kepler\/} stars for which these measures are available in the literature, and show that the stellar masses can be measured with this method to a precision of $\sim$25\%, limited by the surface-gravity precision of the granulation ``flicker" method ($\sim$0.1~dex) and by the parallax uncertainties ($\sim$10\% for the {\it Kepler\/} sample). We explore the impact of expected improvements in the surface gravity determinations---through the application of granulation background fitting and the use of recently published granulation-metallicity relations---and improvements in the parallaxes with the arrival of the {\it Gaia\/} second data release. We show that the application of this methodology to stars that will be observed by {\it TESS\/} should yield radii good to a few percent and masses good to $\approx$10\%. Importantly, the method does not require the presence of an orbiting, eclipsing, or transiting body, nor does it require spatial resolution of the stellar surface. Thus we can anticipate the determination of fundamental, accurate stellar radii and masses for {hundreds of thousands} of bright single stars---across the entire sky and spanning the Hertzsprung-Russell diagram---including those that will ultimately be found to host planets. 
\end{abstract}

\section{Introduction}\label{sec:intro}

Measurements of fundamental physical stellar parameters, especially mass and radius, are paramount to our understanding of stellar evolution. However, at present, different physical prescriptions in stellar evolution models for, e.g., winds, mass-loss, and convective overshoot predict different radii and temperatures for stars of the same mass, age, and metallicity.  Similarly, stars with different elemental abundance ratios will have significant different evolutionary paths in the Hertzsprung-Russell diagram even if they have the same mass and overall metal abundance.   Thus, placing precise constraints on these parameters are critical to constraining the wide range of plausible stellar evolution models. 

One notable problem in stellar astrophysics for which accurate stellar masses and radii are particularly pertinent is the so-called ``radius inflation" of low-mass stars, whose radii have been found in many casess to be significantly larger than model-predicted radii at fixed mass \teff\ by up to 10\% (cf.\ \citealt{Mann2015,Birkby2012}).  To make matters worse, there exists a paucity of isolated M-dwarfs with precisely determined radii in the literature. Moreover, in sparsely populated areas of the Hertzsprung-Russel (HR) diagram---e.g., the Hertzsprung gap, wherein intermediate- and high-mass ($M_{\rm ZAMS} \gtrsim 1.5 \msun$) stars have ceased core hydrogen burning but have not yet ignited hydrogen in their shells---stellar evolution models are poorly constrained.  Thus improving the precision with which we measure the fundamental parameters of the few stars in this regime provides the most promising way of constraining this short-lived phase of stellar evolution. 

A similar issue applies in the case of exoplanet radii and masses, which depend directly on the assumed radii and masses of their host stars. The determination of accurate, empirical masses and radii of planet-hosting stars would in turn enable the accurate, empirical determination of exoplanet radii. 

To date, double-lined eclipsing binaries and stars with angular radii measured interferometrically and distances measured by parallax provide the most robustly determined model-independent stellar radii. The canonical \citet{Torres:2010} sample contains double-lined eclipsing binaries (and $\alpha$ Centauri A and B) with masses and radii good to better than 3\%, but the sample contains only four M dwarfs. \citet{Birkby2012} lists a few dozen M dwarfs in eclipsing binaries or with radii known from interferometry, but the uncertainty in the radii of the stars this sample is as large as $6.4\%$. Interferometry provides radii (via angular diameters) to $\sim$1.5\% for AFG stars \citep{Boyajian2012a} and $\sim$5\% for K and M dwarfs \citep{Boyajian2012b}, but this technique is limited to very bright (and thus nearby) stars.  Among young, low-mass pre-main-sequence stars, there is a severe paucity of benchmark-quality eclipsing binaries, limiting empirical tests of star formation and evolution models \citep[e.g.,][]{Stassun:2014}. Moreover, there is strong evidence that magnetic activity affects the structure of low-mass stars, and can lead to so-called ``radius inflation" of K- and M-dwarfs of up to 10--15\% that has yet to be fully captured in stellar models \citep[see, e.g.,][]{Stassun:2012,Somers:2017}.


A methodology for determining empirical radii of stars using published catalog data has been demonstrated by \citet{StassunGaiaPlanets:2017} for some 500 planet-host stars, in which measurements of stellar bolometric fluxes and temperatures obtained via the available broadband photometry from {\it GALEX\/} to {\it WISE\/} permitted determination of accurate, empirical angular diameters, which, with the {\it Gaia\/} DR1 parallaxes, \citep{GaiaDR1} permitted accurate and empirical measurement of the stellar radii. The improved measurements of the stellar radii permitted an accurate redetermination of the planets' radii. 
In \citet{Stevens:2017}, we extended this methodology to non-planet-hosting stars more generally, again utilizing {\it GALEX\/} through {\it WISE\/} broadband fluxes in order to determine effective temperatures, extinctions, bolometric fluxes, and thus angular radii. We were then able to determine empirical radii for $\sim$125,000 of these stars for which {\it Gaia\/} DR1 parallaxes were available.

For the transiting planet-host star sample analyzed by \citet{StassunGaiaPlanets:2017}, the transit data provide a measure of the stellar density, and thus the stellar mass via the stellar radius. This in turn permitted the transiting planets' masses to be redetermined empirically and accurately. 
Fundamentally, this approach to empirical stellar masses relies---as with eclipsing binary stars---on the orbit and transit of another body about the star. The fundamental stellar mass-radius relationship determined via the gravitational interaction of a star and another body can leave open the question of whether the companion has altered the properties of the star in question (especially in the case of close binary stars). For example, binary stars and close-in star-planet systems can affect one anothers' spin rates and thus activity levels, which can in turn lead to radius inflation and other effects that differ from the basic physics of single-star evolutionary models \citep[see, e.g.,][]{Lopez-Morales:2007,Morales:2008,Privitera:2016}. 

In this paper we seek to develop a pathway to empirical, accurate masses of single stars. The approach makes use of the fact that an individual star's surface gravity is accurately and independently encoded in the amplitude of its granulation-driven brightness variations \citep[e.g.,][]{Bastien:2013,Corsaro14,Corsaro15,Kallinger16,Bastien:2016}---variations which can be measured with precise light curve data such as will soon become available for bright stars across the sky with the {\it Transiting Exoplanet Survey Satellite (TESS)} \citep{Ricker:2015} and, later, {\it PLATO} \citep{Rauer:2014}. Combined with an accurate stellar radius determined independently via the broadband SED and the {\it Gaia\/} parallax as described above, the stellar mass follows directly. 

Of course, for stars found to possess planets, such accurate, empirical stellar masses and radii will permit determination of the exoplanet radii and masses also. Indeed, applying this approach to targets that will be observed by the upcoming {\it TESS\/} and {\it PLATO\/} missions could help to optimize the search for small transiting planets \citep[see][]{StassunTESS:2014,Campante:2016}.
Most importantly, empirical stellar masses and radii determined in this fashion for single stars---without stellar or planetary companions---should enable progress on a number problems in stellar astrophysics, including radius inflation in low-mass stars, and will provide a large set of fundamental testbeds for basic stellar evolution theory of single stars. 

In Section~\ref{sec:methods} we describe our methodology and the extant data that we utilize to demonstrate the method. Section~\ref{sec:results} presents our results, including estimates for the expected precision with which stellar masses may be measured and the limits of applicability. In Section~\ref{sec:discussion} we discuss the likely number of {\it TESS\/} stars likely to yield accurate stellar mass determinations, some example applications of the stellar masses so determined, as well as some caveats, potential sources of systematic error, and how these might be mitigated. We conclude with a summary of our conclusions in Section~\ref{sec:summary}.

\section{Data and Methods}\label{sec:methods}

\subsection{Data from the literature}

In order to demonstrate our approach in a manner that is as similar as possible to what we expect from the upcoming {\it TESS\/} and {\it Gaia\/} datasets, we draw our sample data from two recent studies of large numbers of {\it Kepler\/} stars. In particular, we take as ``ground truth" the asteroseismically determined stellar masses (\mstar) and radii (\rstar), and spectroscopically determined stellar effective temperatures (\teff), from \citet{Huber:2017}. Those authors also report stellar bolometric fluxes (\fbol) and angular radii ($\Theta$) measured via the broadband spectral energy distribution (SED) method laid out in \citet{StassunGaiaEB:2016} and \citet{StassunGaiaPlanets:2017}. Finally, we take the stellar surface gravities (\logg) for these stars as determined via the granulation ``flicker" method from \citet{Bastien:2016}.  
Together, these sources provide a sample of \nsample\ stars for demonstration of the methodology explored in this work.

\subsection{Summary of methodology} 

\subsubsection{Stellar radius via spectral energy distributions}

At the heart of this study is the basic methodology laid out in \citet{StassunGaiaEB:2016} and \citet{StassunGaiaPlanets:2017}, in which a star's angular radius, $\Theta$, can be determined empirically through the stellar bolometric flux, \fbol, and effective temperature, \teff, according to
\begin{equation}\label{eq:frt}
\Theta = ( F_{\rm bol} / \sigma_{\rm SB} T_{\rm eff}^4 )^{1/2},
\end{equation}
\noindent where $\sigma_{\rm SB}$ is the Stefan-Boltzmann constant.

\fbol\ is determined empirically
by fitting stellar atmosphere models to the star's observed SED, assembled from archival broadband photometry over as large a span of wavelength as possible, preferably from the ultraviolet to the mid-infrared. 
\teff\ is ideally taken from spectroscopic determinations when available, in which case the determination of \fbol\ from the SED involves only an estimate of the extinction, $A_V$, and an overall normaliztion as free parameters. 

\begin{figure}[!ht]
\begin{center}
\includegraphics[width=\linewidth,trim=0pt 0pt 0pt 50pt,clip]{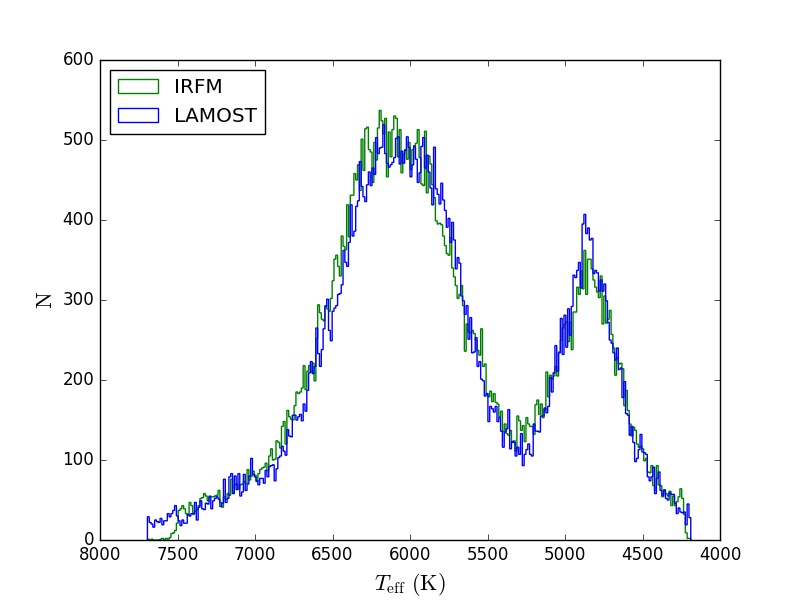}
\includegraphics[width=\linewidth,trim=0pt 0pt 0pt 50pt,clip]{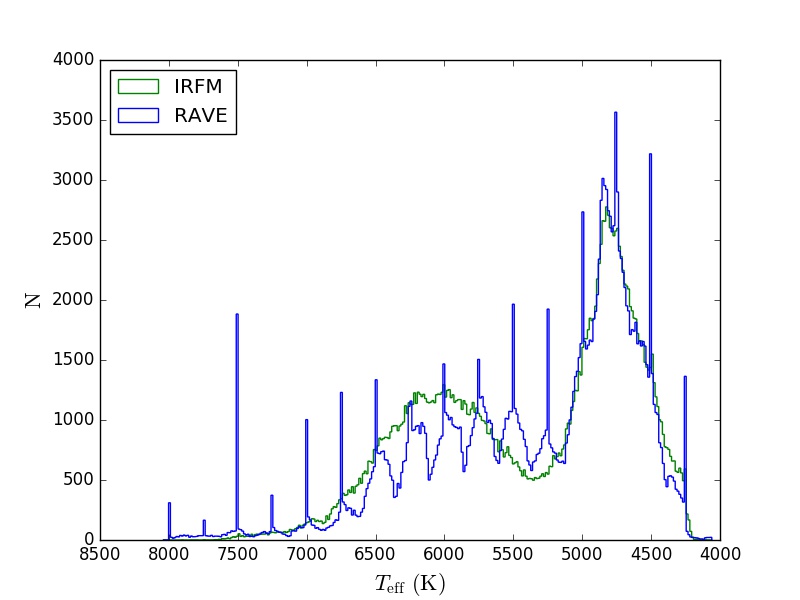}
\includegraphics[width=\linewidth,trim=0pt 0pt 0pt 50pt,clip]{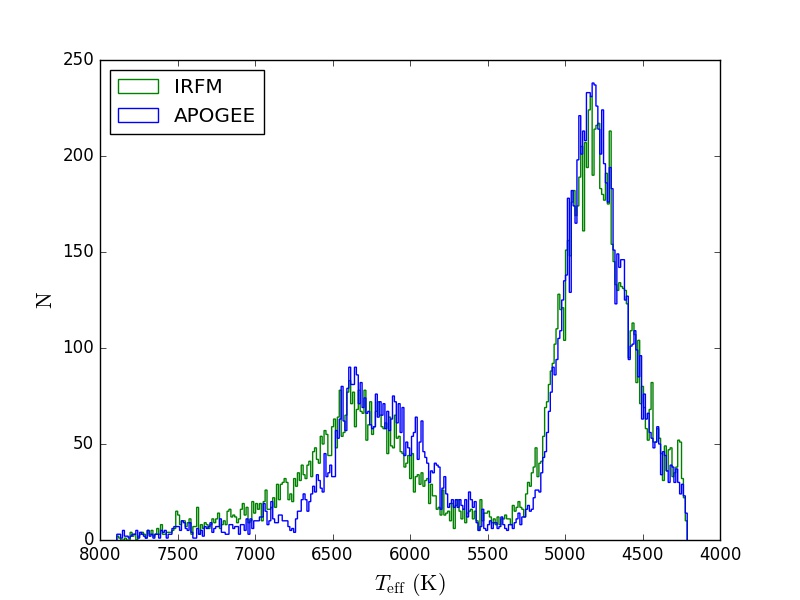}
\end{center}
\vspace{-0.15in}
\caption{\label{fig:teffcomp2} Spectroscopic versus best-fit SED-based \teff\ (labelled IRFM in the plots) using stars in LAMOST (\emph{top}), RAVE (\emph{middle}), and APOGEE (\emph{bottom}) catalogs as checks, showing that our procedures are able to recover the spectroscopically determined \teff, generally to within $\sim$150~K. The peaks in the RAVE histogram correspond to the grid resolution of synthetic spectra used by the RAVE pipeline. Reproduced from \citet{Stevens:2017}.}
\end{figure}

If \teff\ is not available from spectroscopic determinations, then \teff\ may also be determined from the SED as an additional fit parameter, as we showed in \citet{Stevens:2017}. 
Figure~\ref{fig:teffcomp2} \citep[from][]{Stevens:2017} shows the performance of our procedures when we also determine \teff\ as part of the SED fitting process (here using LAMOST, RAVE, and APOGEE spectroscopic \teff\ as checks). 
Our SED-based procedure recovers the spectroscopically determined \teff, generally to within $\sim$150~K. It does appear that our method infers an excess of stars with $\teff > 7,000$~K (Fig.~\ref{fig:teffcomp2}, bottom), suggesting somewhat larger \teff\ uncertainties of $\sim$250~K for stars hotter than about 7,000~K.

\begin{figure}[!ht]
    \centering
    \includegraphics[width=\linewidth,trim=70 70 70 50,clip]{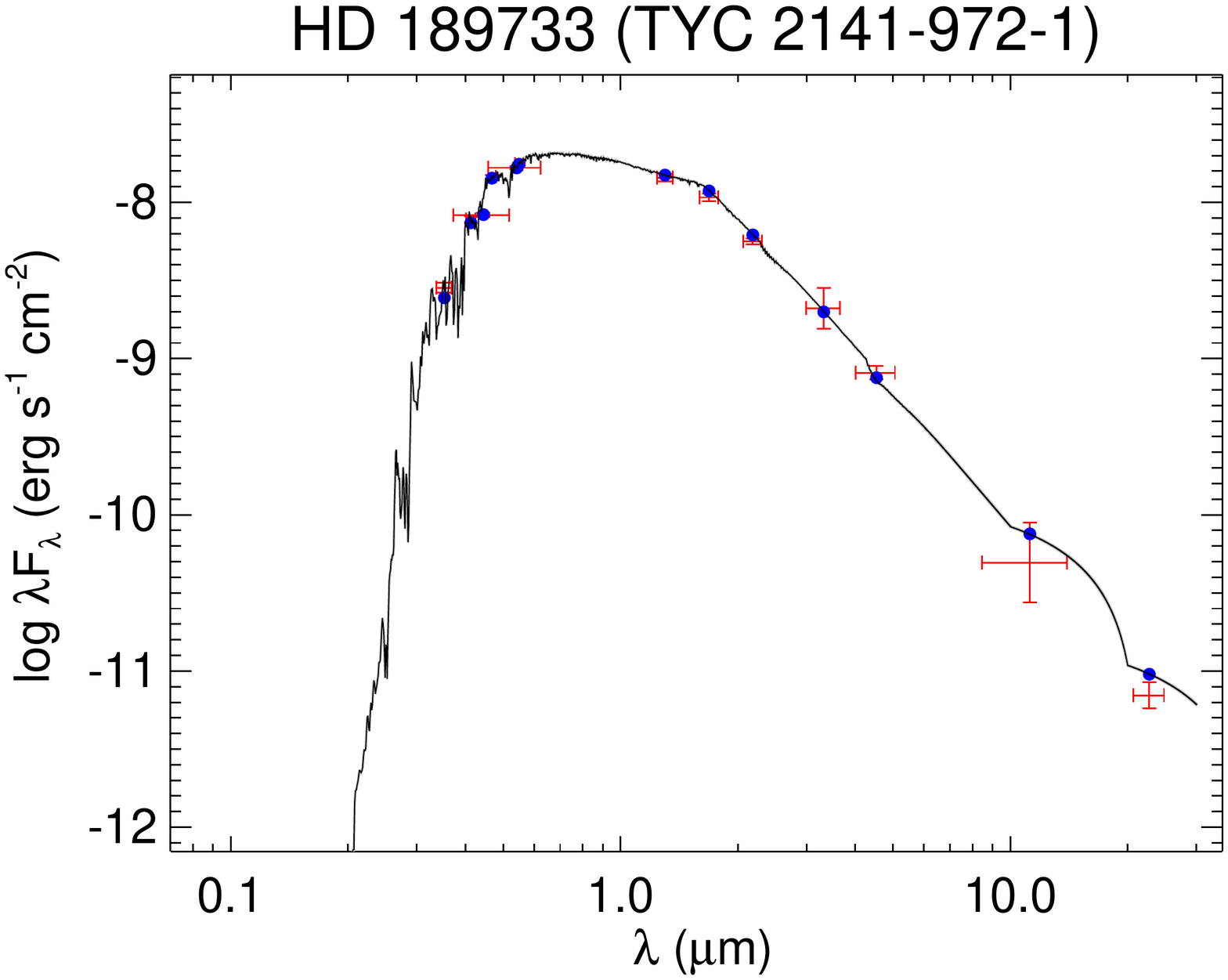}
    \includegraphics[width=\linewidth,trim=70 70 70 50,clip]{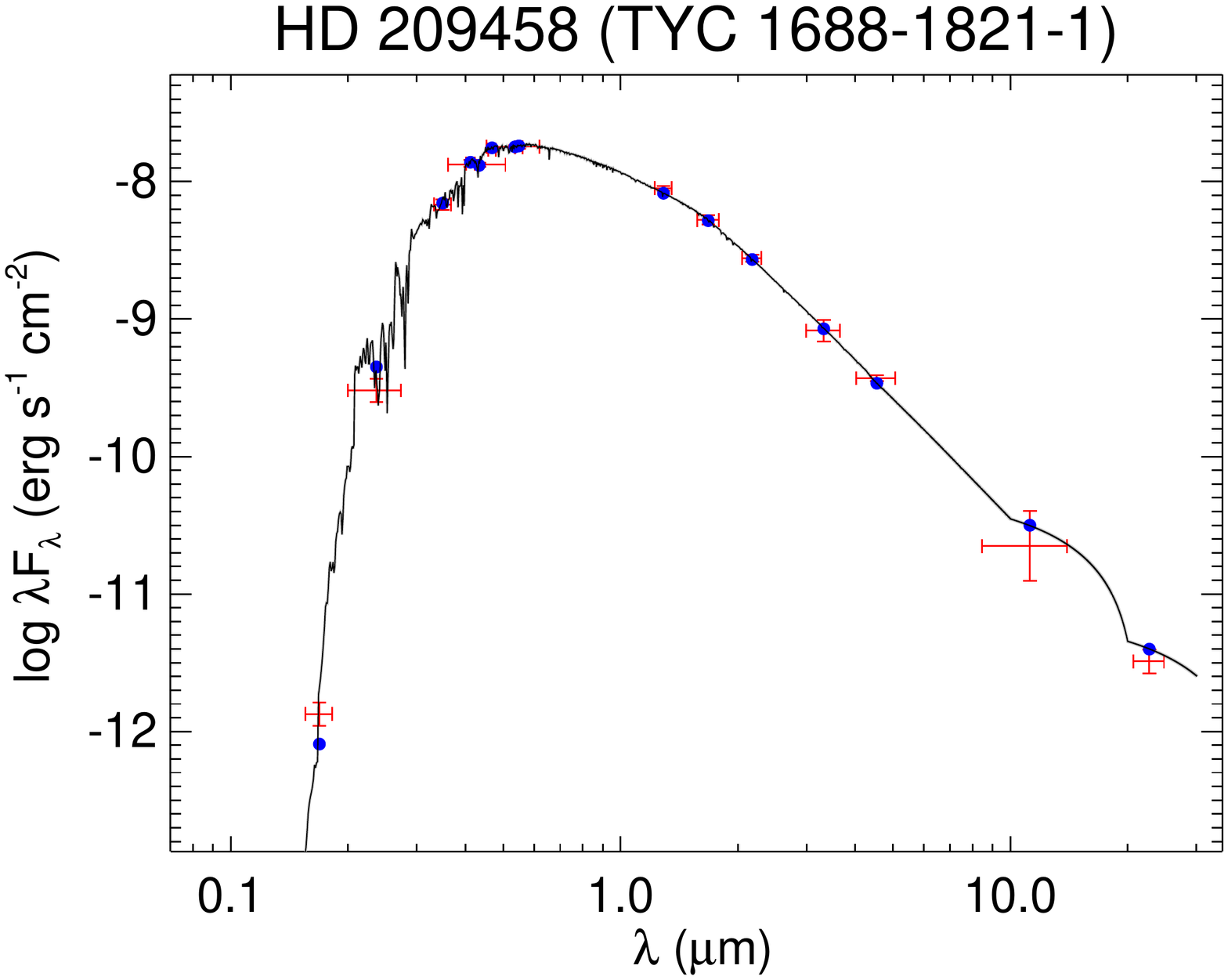}
    \caption{SED fits for the stars HD~189733 and HD~209458, for which interferometric angular radii have been reported \citep{Boyajian:2015} as a check on the $\Theta$ and \fbol\ values derived via our methodology. Each panel shows the observed fluxes from {\it GALEX\/} to {\it WISE\/} vs.\ wavelength (in \micron) as red error bars, where the vertical error bar represents the measurement uncertainty and the horizontal ``error" bar represents the width of the passband. Also in each figure is the fitted SED model including extinction, on which is shown the model passband fluxes as blue dots. 
    The two SED fits have goodness-of-fit $\chi_\nu^2$ of 1.65 and 1.67, respectively. The $\Theta$ and \fbol\ comparisons are presented in Table~\ref{tab:boyajian}. Reproduced from \citet{StassunGaiaPlanets:2017}.}
    \label{fig:boyajian}
\end{figure}

For the purposes of this demonstration study we utilize only \teff\ determined spectroscopically \citep{Huber:2017}.
As a demonstration of our SED fitting approach, in \citet{StassunGaiaPlanets:2017} we applied our procedures to the interferometrically observed planet-hosting stars HD~189733 and HD~209458 reported by \citet{Boyajian:2015}. Our SED fits are reproduced in Figure~\ref{fig:boyajian} and the $\Theta$ and \fbol\ values directly measured by those authors versus those derived in this work are compared in Table~\ref{tab:boyajian}, where the agreement is found to be excellent and within the uncertainties. 

\begin{deluxetable*}{llcc}
\tablecolumns{4}
\tablewidth{0pt}
\tablecaption{Comparison of stellar angular diameters ($2\times\Theta$) and \fbol\ for stars with interferometric measurements from \citet{Boyajian:2015} versus the SED based determinations from \citet{StassunGaiaPlanets:2017}.
\label{tab:boyajian}}
\tablehead{\colhead{} & \colhead{} & \colhead{\citet{Boyajian:2015}} & \colhead{\citet{StassunGaiaPlanets:2017}}} 
\startdata
HD~189733 & $2\times\Theta$ (mas) & 0.3848$\pm$0.0055 & 0.391$\pm$0.008 \\
 & \fbol\ ($10^{-8}$ erg s$^{-1}$ cm$^{-2}$) & 2.785$\pm$0.058 & 2.87$\pm$0.06 \\
HD~209458 & $2\times\Theta$ (mas) & 0.2254$\pm$0.0072 & 0.225$\pm$0.008 \\
 & \fbol\ ($10^{-8}$ erg s$^{-1}$ cm$^{-2}$) & 2.331$\pm$0.051 & 2.33$\pm$0.05 \\
\enddata
\end{deluxetable*}

The examples in Fig.~\ref{fig:boyajian} represent cases where the stellar \teff\ was drawn from spectroscopic determinations \citep[via the PASTEL catalog;][]{Soubiran:2016}. 
In this study, the \teff\ values we adopt are also spectroscopic, determined via the ASPCAP pipeline applied to the APOKASC {\it APOGEE-2\/} high-resolution, near-infrared spectra of the {\it Kepler\/} field. 
For future applications to the {\it TESS\/} stars, 
the {\it TESS\/} Input Catalog (TIC) will provide spectroscopic \teff\ a large fraction of the target stars as well as photometrically estimated \teff\ for the vast majority of other targets \citep{StassunTIC:2017}.

\begin{figure}[!ht]
\centering
\includegraphics[width=\linewidth,trim=10 10 10 50,clip]{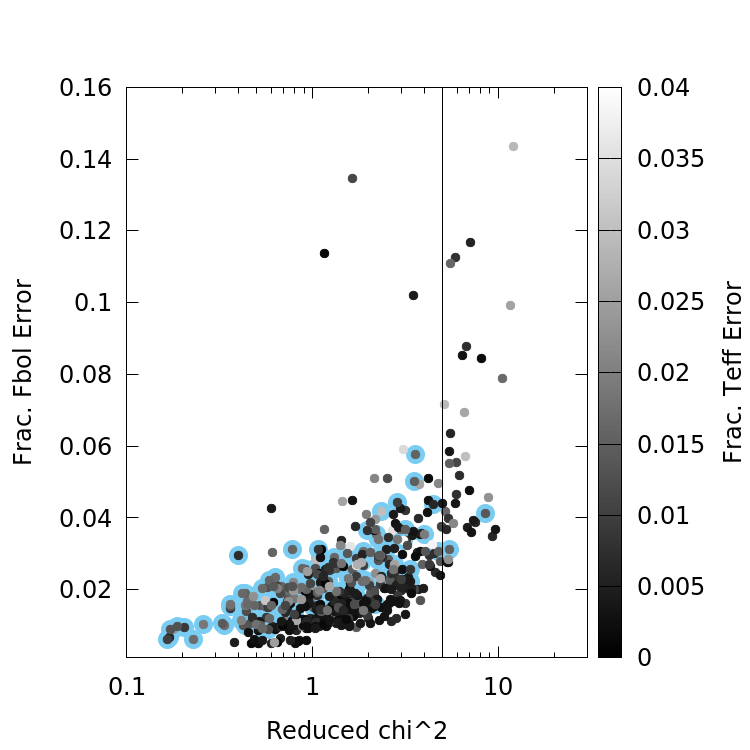}
\caption{Fractional uncertainty on \fbol\ from the SED fitting procedure as a function of $\chi_\nu^2$ and of \teff\ uncertainty. The vertical line represents the cutoff of $\chi_\nu^2 \le 5$ for which the uncertainty on \fbol\ is at most 6\% for most stars, thus permitting a determination of \rstar\ to $\approx$3\%. Points with blue haloes represent stars with transiting planets. 
Reproduced from \citet{StassunGaiaPlanets:2017}.
\label{fig:fbol_vs_chi2}}
\end{figure}

As demonstrated in \citet{StassunGaiaPlanets:2017}, with this wavelength coverage for the constructed SEDs, the resulting \fbol\ are generally determined with an accuracy of a few percent when \teff\ is known spectroscopically, though the uncertainty can be as large as $\sim$10\% when \teff\ is obtained as part of the SED fitting \citep{Stevens:2017}. 
Figure~\ref{fig:fbol_vs_chi2} shows the fractional \fbol\ uncertainty for the sample from \citet{StassunGaiaPlanets:2017} as a function of the goodness of the SED fit and of the uncertainty on \teff. For stars with \teff\ uncertainties of $\lesssim$1\%, the \fbol\ uncertainty is dominated by the SED goodness-of-fit. With the exception of a few outliers, it was shown that one can achieve an uncertainty on \fbol\ of at most 6\% for $\chi_\nu^2 \le 5$,
with 95\% of the sample having an \fbol\ uncertainty of less than 5\%. 
{As discussed in \citet{StassunGaiaPlanets:2017}, outliers in Fig.~\ref{fig:fbol_vs_chi2} likely represent the small fraction of stars that are unresolved binaries comprising stellar components that simultaneously have sufficiently different \teff\ and sufficiently comparable brightness; such binaries are easily screened out via the SED $\chi_\nu^2$ metric.}

For the purposes of this demonstration study, we require $\chi_\nu^2 < 10$ and the relative uncertainty on the parallax, $\sigma_\pi / \pi$, to be at most 20\% \citep[see, e.g.,][for a discussion]{BailerJones:2015}. 
This leaves a final study sample of \nsamplefin\ stars.

\subsubsection{Stellar surface gravity via granulation-driven brightness variations}

The granulation-based \logg\ measurements that we use from \citet{Bastien:2016} are based on the ``flicker" methodology of \citet{Bastien:2013}. That method uses a simple measure of the r.m.s.\ variations of the {\it Kepler\/} light curve on an 8-hr timescale (\flick), representing the meso-granulation driven brightness fluctuations of the stellar photosphere. Importantly, as demonstrated by \citet{Bastien:2013,Bastien:2016}, the \flick\ amplitude is measureable even if the instrumental shot noise is up to $\sim$5 times larger than the \flick\ signal itself, so long as the shot noise as a function of stellar apparent magnitude can be well characterized. For example, in the {\it Kepler\/} sample analyzed by \citet{Bastien:2016}, the \flick\ amplitude of $\sim$15~p.p.m.\ for solar-type dwarfs could be reliably measured in stars as faint as 13th magnitude in the {\it Kepler\/} bandpass, for which the typical shot noise was $\sim$75~p.p.m. As described by \citet{Bastien:2013,Bastien:2016}, removing the shot noise in quadrature from the directly measured r.m.s.\ allows the \flick\ amplitude as small as $\sim$20\% of the total r.m.s.\ to be measured with sufficient precision to permit the stellar \logg\ to be determined with a typical precision of $\sim$0.1~dex. 

The granulation properties can also be extracted from the so-called ``background" signal in the stellar power spectrum, i.e., the Fourier transform of the light curve from the time domain to the frequency domain. 
This technique was originally proposed by \cite{Harvey85} as applied to the Sun, and is now widely adopted for the analysis of stars observed with {\it Kepler\/} \citep[e.g.][]{Mathur11,Kallinger14}. It consists of modeling
the granulation signal in the power spectrum through its individual components, namely that of granulation, the instrumental photon noise, as well as possible acoustic-driven oscillations. The fitting process is usually performed by means of Monte Carlo Bayesian approaches to better sample the possible correlations arising among the free parameters of the background model
\citep{Kallinger14,Corsaro14,Corsaro15}.

The granulation signal in the power spectrum is modeled using two super-Lorentzian profiles\footnote{A super-Lorentzian profile is defined as a Lorentzian profile with a varying exponent, namely an exponent that is not necessarily equal to 2. Such a super-Lorentzian profile is used to model the characteristic granulation-driven signal in the Fourier domain of a light curve. A Bayesian model comparison performed by \cite{Kallinger14} on a large sample of stars (about 600) observed with NASA Kepler has shown that the most likely exponent of the super-Lorentzian profile is 4. This was also adopted by \citet{Corsaro15} in the asteroseismic study of a sample of red giant stars, and later on used by \citet{Corsaro:2017} for detecting the metallicity effect on stellar granulation.}, one corresponding to the time-scale of the actual granulation and another to that of the meso-granulation, the latter representing a reorganization of the granulation phenomenon at larger spatial scales and longer temporal scales \citet{Corsaro:2017}. Each of these components is defined by two parameters, the amplitude of the signal ($a_{\rm gran}$ for the granulation and $a_{\rm meso}$ for the meso-granulation) and the characteristic frequency ($b_{\rm gran}$ and $b_{\rm meso}$, respectively). As shown by \citet{Kallinger14} and \citet{Corsaro:2017}, the granulation and meso-granulation parameters scale linearly with one another, implying that one need only measure one or the other to fully infer the granulation properties of the star. The characteristic frequencies of this signal are tightly related to the surface gravity of the star since $b_{\rm meso}$ $\propto b_{\rm gran}$ $\propto g / \sqrt{T_{\rm eff}}$ \citep{Brown91}. This means that $g$ can be measured from either $b_{\rm gran}$ or $b_{\rm meso}$ in the stellar power spectrum. 

This granulation background method is typically used as the preliminary step in performing the traditional asteroseismic ``peak bagging" analysis \citep[e.g.][]{Handberg:2011,Corsaro15}, in which individual stellar oscillation frequency peaks are fitted in the power spectrum. The traditional seismic approach remains the preferred method when there is sufficient signal to enable such fine analysis, because in general it yields the most accurate and precise stellar parameters. It does, however, in general require brighter stars that have been observed for enough time, often on the order of several months, to allow resolving the individual modes of oscillation. With {\it TESS}, it is estimated that a few hundred planet-hosting red giants and subgiants (and some F dwarfs) will be amenable to seismic analysis \citep{Campante:2016}.

The background modeling technique has been shown to reach about 4\% precision in $g$ using the full set of observations from {\it Kepler\/} \citep{Kallinger16,Corsaro:2017}. Through a Bayesian fitting of the background properties 
and a detailed Bayesian model comparison, \citet{Corsaro:2017} has recently shown that stellar mass and metallicity play a significant role in changing the parameters that define the granulation-related signal in a sample of cluster red giant stars observed with {\it Kepler}. In particular, the authors detected a 20--25\% decrease in $b_{\rm meso}$ with an increase in mass of $\sim$0.5~\msun, and a 30--35\% decrease in $b_{\rm meso}$ with an increase in metallicity of $\sim$0.3~dex. This also implies that the accuracy in $g$ from the background modeling can be further improved by taking into account the mass and metallicity of the stars using, e.g., the empirical relations of \citet{Corsaro:2017}.

\section{Results}\label{sec:results}

In this section we summarize the results of our methodology to determine empirical stellar masses in three steps. First, we demonstrate the granulation-based \logg\ precision that may be expected from {\it TESS\/} light curves. Second, we demonstrate the precision on \rstar\ that may be expected from SED-based \fbol\ together with {\it Gaia\/} parallaxes. Then we demonstrate the precision on \mstar\ that may be expected via the combination of \logg\ and \rstar\ from the first two steps.

\subsection{Expected Precision of Surface Gravity}\label{subsec:logg}

\subsubsection{Flicker}

\begin{figure*}[!ht]
    \centering
    \includegraphics[trim=75 510 60 90,clip,width=\linewidth]{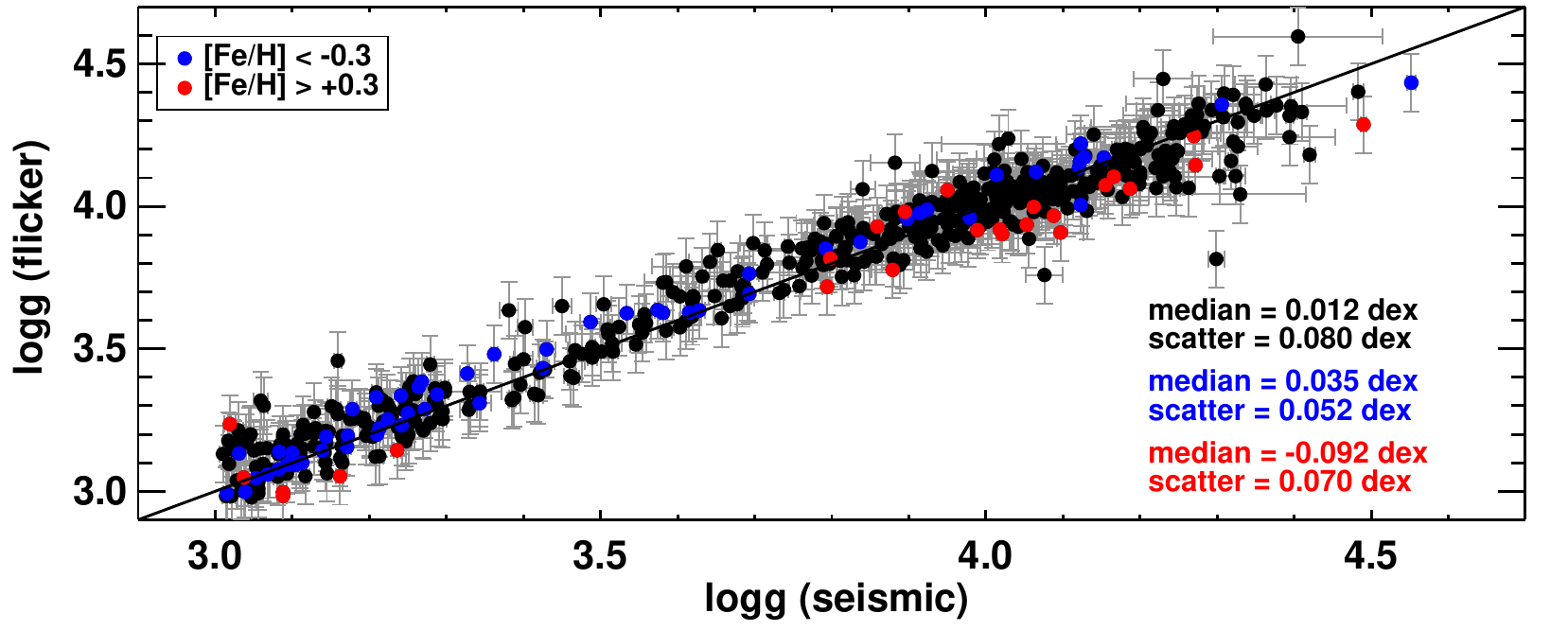}
    \caption{Comparison of \logg\ obtained via granulation ``flicker" \citep{Bastien:2016} versus those obtained asteroseismically \citep{Huber:2017}. The overall agreement has an r.m.s.\ scatter of 0.08~dex. Sub-dividing the sample into metal-rich and metal-poor improves the agreement to 0.05--0.07~dex, as suggested by \citep{Corsaro:2017}.}
    \label{fig:flicker}
\end{figure*}

We begin by verifying that the granulation-based \logg\ fundamentally agrees with that obtained via asteroseismology. To do this, we show in Figure~\ref{fig:flicker} the comparison of the \flick-based \logg\ from \citet{Bastien:2016} versus the seismic \logg\ from \citet{Huber:2017}. 
The agreement is excellent, with an overall offset of 0.01~dex and r.m.s.\ scatter of 0.08~dex. 
This is of course not surprising, since the \flick\ method was originally calibrated on asteroseismic samples \citep{Bastien:2013}. 

At the same time, this comparison also corroborates the finding by \citet{Corsaro:2017} that the granulation-based \logg\ determination involves a metallicity dependence. 
In Figure~\ref{fig:flicker} we see that by subdividing the sample into a metal-rich subset and a metal-poor subset, the agreement between the \flick-based \logg\ and the asteroseismic \logg\ improves to as good as 0.05~dex. 
While we do not implement any metallicity corrections in this demonstration study, in future work we expect that using, e.g., the empirical metallicity correction of \citet{Corsaro:2017} should improve the accuracy of granulation-based \logg. 

The {\it TESS\/} light curves are expected to have a systematic noise floor of {that could be as large as} 60~ppm \citep{Ricker:2015}, which would dominate the error budget for most bright stars. Meanwhile, the \flick\ amplitude of solar-type stars is $\approx$15~ppm \citep{Bastien:2013}. As noted above, the \flick\ amplitude was found to be measurable in the {\it Kepler\/} light curves even down to $\sim$20\% of the noise. Thus the solar-type \flick\ signal is measurable for noise levels as high as $\sim$75~ppm. In addition, the \flick\ method involves averaging the light curve on 8-hr timescale, or 16 frames for the 30-min FFI data. For {\it TESS\/} FFI data, therefore, the 75~ppm noise limit corresponds to a 300~ppm per-image noise limit, or approximately 10.5~mag in the {\it TESS\/} bandpass.

This is a significantly brighter limit than was the case for {\it Kepler}; the \flick\ signal was extracted successfully for {\it Kepler\/} stars as faint as 14~mag \citep{Bastien:2016}. 
We discuss the implications for the accessible {\it TESS\/} target sample in Section~\ref{sec:discussion}.

\subsubsection{Granulation background modeling}

As noted in Section~\ref{sec:methods}, the granulation signal has also been shown to be measurable via modeling of the so-called ``granulation background" in Fourier space, leading to the measurement of \logg\ with considerably improved precision over the \flick\ method. 
Here we present simulated results of such an approach in the {\it TESS\/} context. 

\begin{figure*}[!ht]
    \centering
    \includegraphics[width=0.5\linewidth]{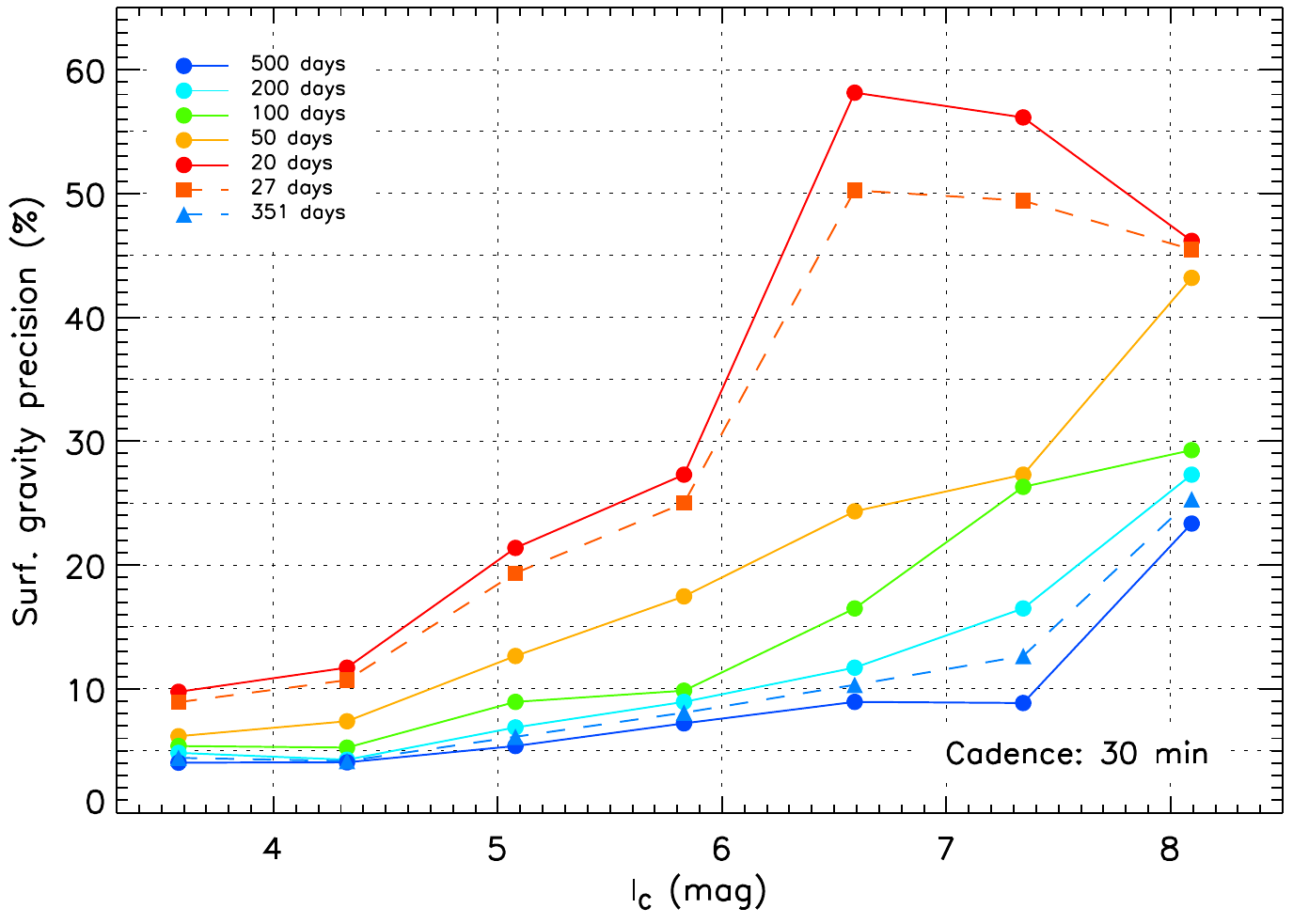}\includegraphics[width=0.5\linewidth]{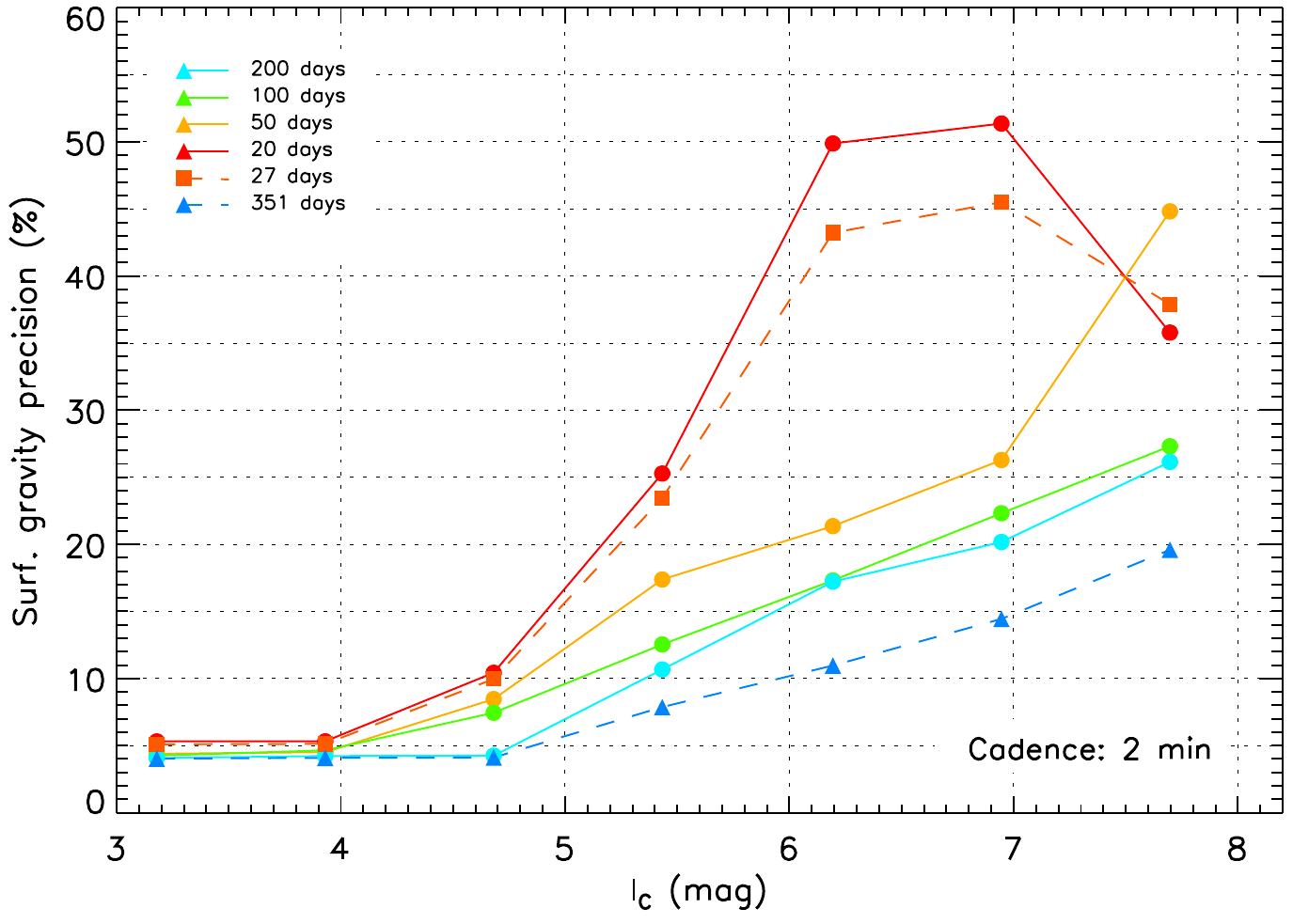}
    \includegraphics[width=0.5\linewidth]{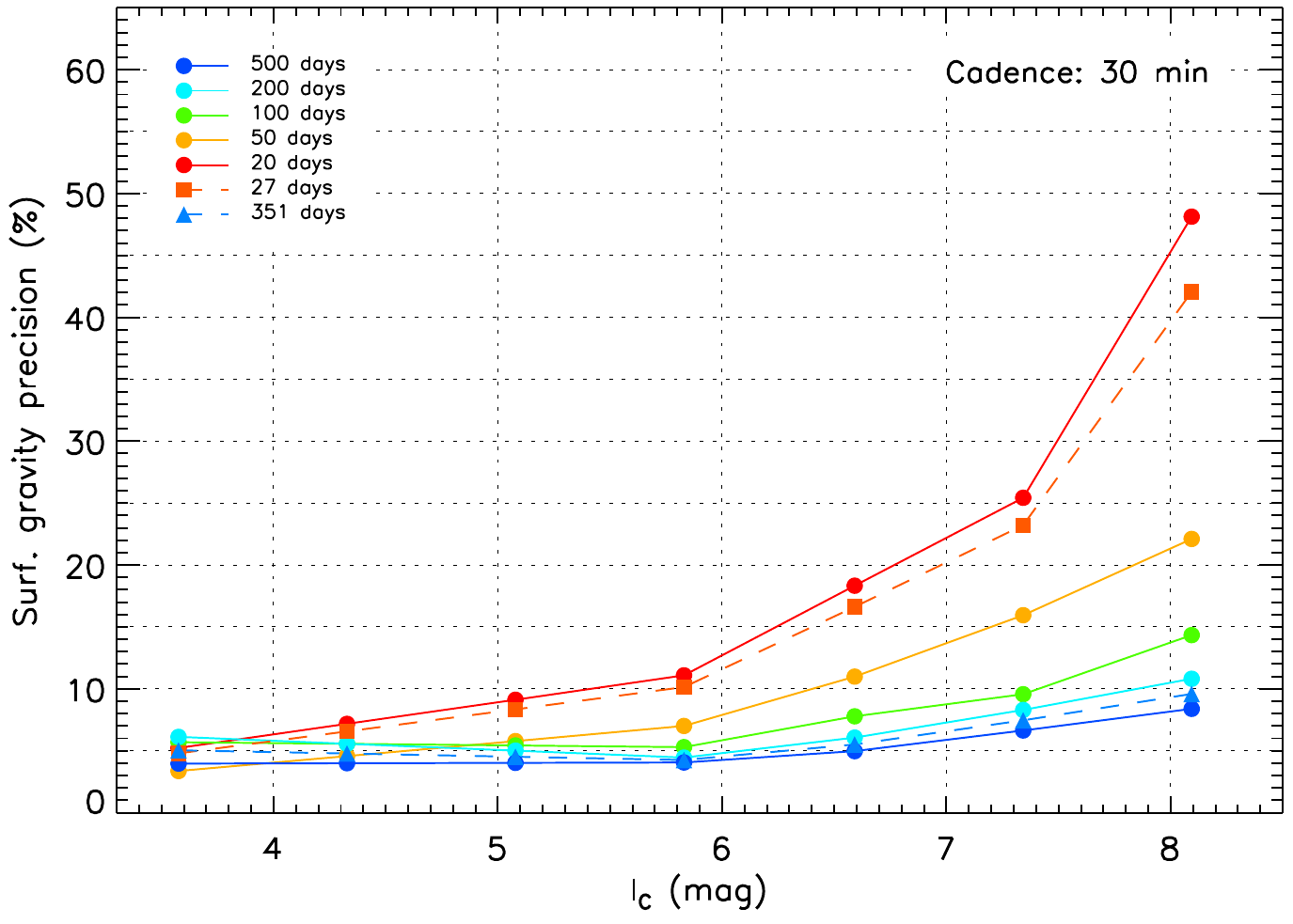}\includegraphics[width=0.5\linewidth]{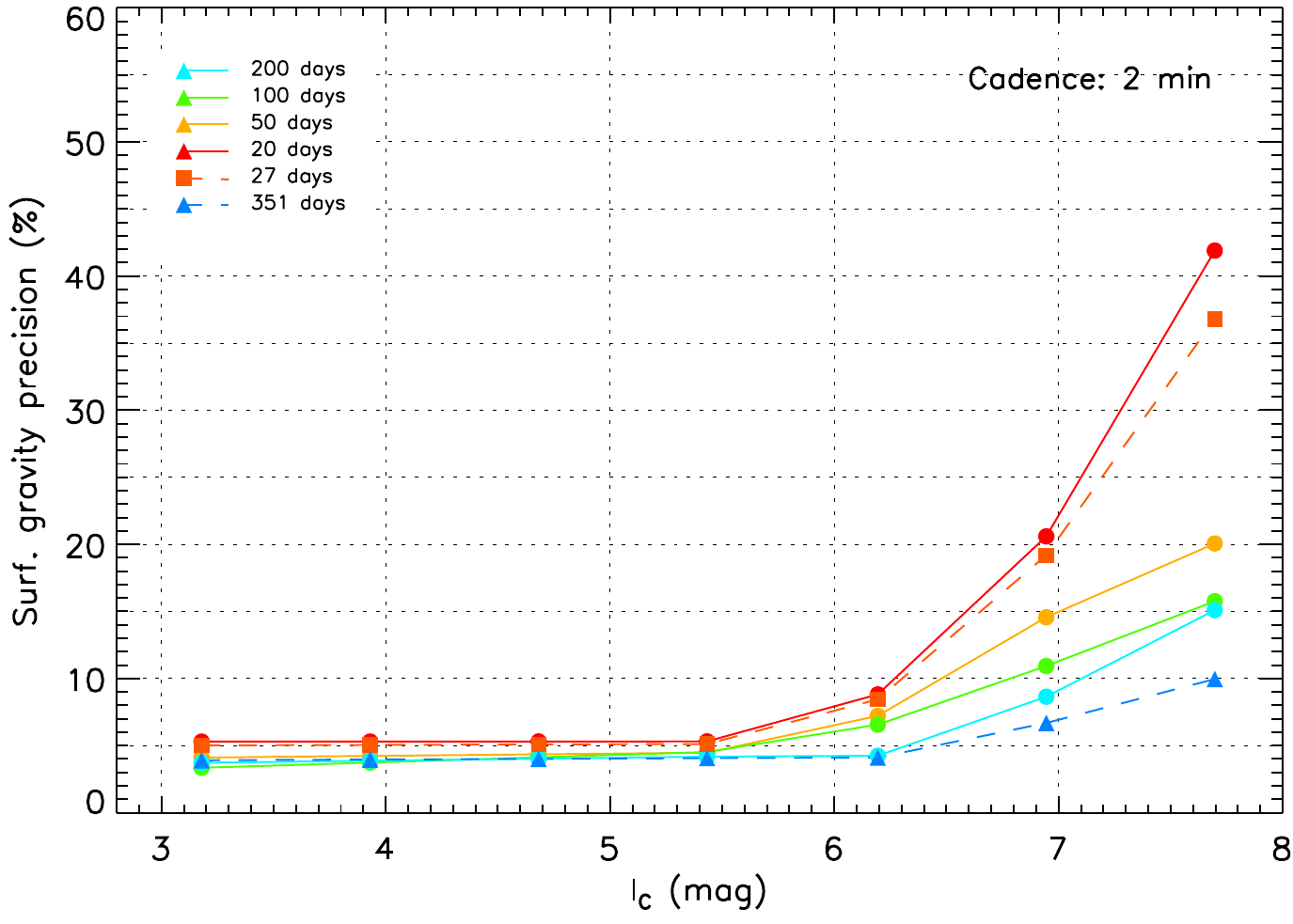}    
    \caption{Simulated precision on stellar surface gravity, $g$, from the granulation background modeling technique \citep[e.g.,][]{Kallinger16,Corsaro:2017} applied to {\it TESS\/} light curves. (Top row:) Results for 30-min (left) and 2-min (right) cadence light curves of various durations, requiring the granulation signal to be at least as large as the photometric noise. The dashed lines correspond to cases with durations typical of the {\it TESS\/} instrument. (Bottom row:) Same as top row, but now requiring the granulation signal to be only 20\% as large as the photometric noise \citep[see][]{Bastien:2013,Bastien:2016}.}
    \label{fig:corsaro}
\end{figure*}

Figure~\ref{fig:corsaro} (top row) presents the precision expected for $g$, depending on the light curve cadence and on the total light curve duration (the precision of the background modeling method is sensitive to these parameters because it is fundamentally based on fitting the Fourier spectrum).
The figure incorporates the results from \citet{Kallinger16}, which were based on the {\it Kepler\/} 30-min and 1-min cadences. Those authors extracted the precision on $g$ by using real {\it Kepler\/} data and by adding noise to the light curve and/or degrading the total observing time of the dataset. We have used those data here because their methodology closely resembles that which we used in our analysis of the metallicity effect on the granulation amplitudes \citep{Corsaro:2017}. In particular, their $\tau_{\rm ACF}$ parameter is comparable to the $b_{\rm meso}$ parameter from the fits presented in \citet{Corsaro:2017}, yielding the same precision. Since both analyses depend on the level of signal-to-noise and on the frequency resolution of the power spectrum in the same way, we can map the \citet{Kallinger16} {\it Kepler\/} results onto the simulated {\it TESS\/} data.

The short-cadence case represents the convection-driven oscillations of a solar-type main-sequence star, having $\nu_{\rm max}$ $\simeq 1000\,\mu$Hz, and can be considered appropriate also for stars up to the subgiant regime ($\nu_{\rm max} \gtrsim 300$~$\mu$Hz). The long-cadence case represents instead a red giant and therefore applies for stars with $\nu_{\rm max} < 300$~$\mu$Hz (typically 50--200~$\mu$Hz).
We converted the {\it Kepler\/} magnitudes from the simulations \citet{Kallinger16} into Cousins $I$-band ($I_C$) as comparable to the {\it TESS\/} instrument by taking into account the $\sim$10 times higher noise level expected in the power spectra for a star observed by {\it TESS}. This translates into a shift in magnitude (for a given signal-to-noise ratio) to 5~mag brighter for {\it TESS\/} targets.



In order to achieve a precision that is better than what is achievable with the \flick\ method, we can require a precision of $\sim$0.04~dex in \logg\ or $\sim$10\% in $g$. To satisfy this condition we require 
$I_C < 4.7$ for 30-min cadence, and $I_C < 4.3$ for 2-min cadence, for a 27-day observation. Similarly we have
$I_C < 6.6$ for 30-min cadence, and $I_C < 5.4$ for 2-min cadence, for a 351-day observation. 
Note that as the apparent magnitude increases, the 30-min cadence precision shows a less steep rise compared to the 2-min cadence; this is the result of the increase in the amplitude of the granulation signal with the evolution of the star (the simulated long cadence case is a red giant). 

As noted by \citet{Kallinger16}, there is no particular limitation on the detectability of the granulation background signal, assuming that the timescales stay within the Nyquist frequency imposed by the cadence.
However, for simplicity the simulations of \citet{Corsaro:2017} required the granulation amplitude to be at least as large as the photometric noise. 
As noted above, \citet{Bastien:2013,Bastien:2016} found that the granulation signal is measurable in practice down to $\sim$20\% of the photometric noise, which would extend the reach and precision of the granulation background modeling approach to fainter {\it TESS\/} stars. 

The effect of this is shown in Figure~\ref{fig:corsaro} (bottom row), again for both the 30-min and 2-min cadences and for a range of light-curve time baselines. Now it becomes possible to measure $g$ with 10\% precision down to $\sim$6~mag for the 30-min cadence and down to $\sim$6.5~mag for the 2-min cadence. 
{It is also possible to measure $g$ with a precision comparable to that of the \flick\ method ($\sim$20\%) down to $\approx$7~mag for dwarfs/subgiants in the 2-min cadence and for red giants in the 30-min cadence.}
{For comparison, it is estimated that a full asteroseismic analysis can be done for subgiants (and some dwarf stars) for stars brighter than $\sim$5~mag \citep[see, e.g.,][]{Campante:2016}.}
We discuss the implications for the accessible {\it TESS\/} target sample in Section~\ref{sec:discussion}.

\subsection{Expected Precision of Stellar Radii}\label{subsec:radius}

Next we consider the expected precision on \rstar\ that may be achieved through the method of \fbol\ via broadband SED fitting together with the parallax from {\it Gaia}. Here we utilize the same demonstration sample as above, comparing the \rstar\ inferred from the SED+parallax against the \rstar\ obtained asteroseismically. 

\begin{figure*}[!ht]
    \centering
    \includegraphics[trim=75 75 60 300,clip,width=0.95\linewidth]{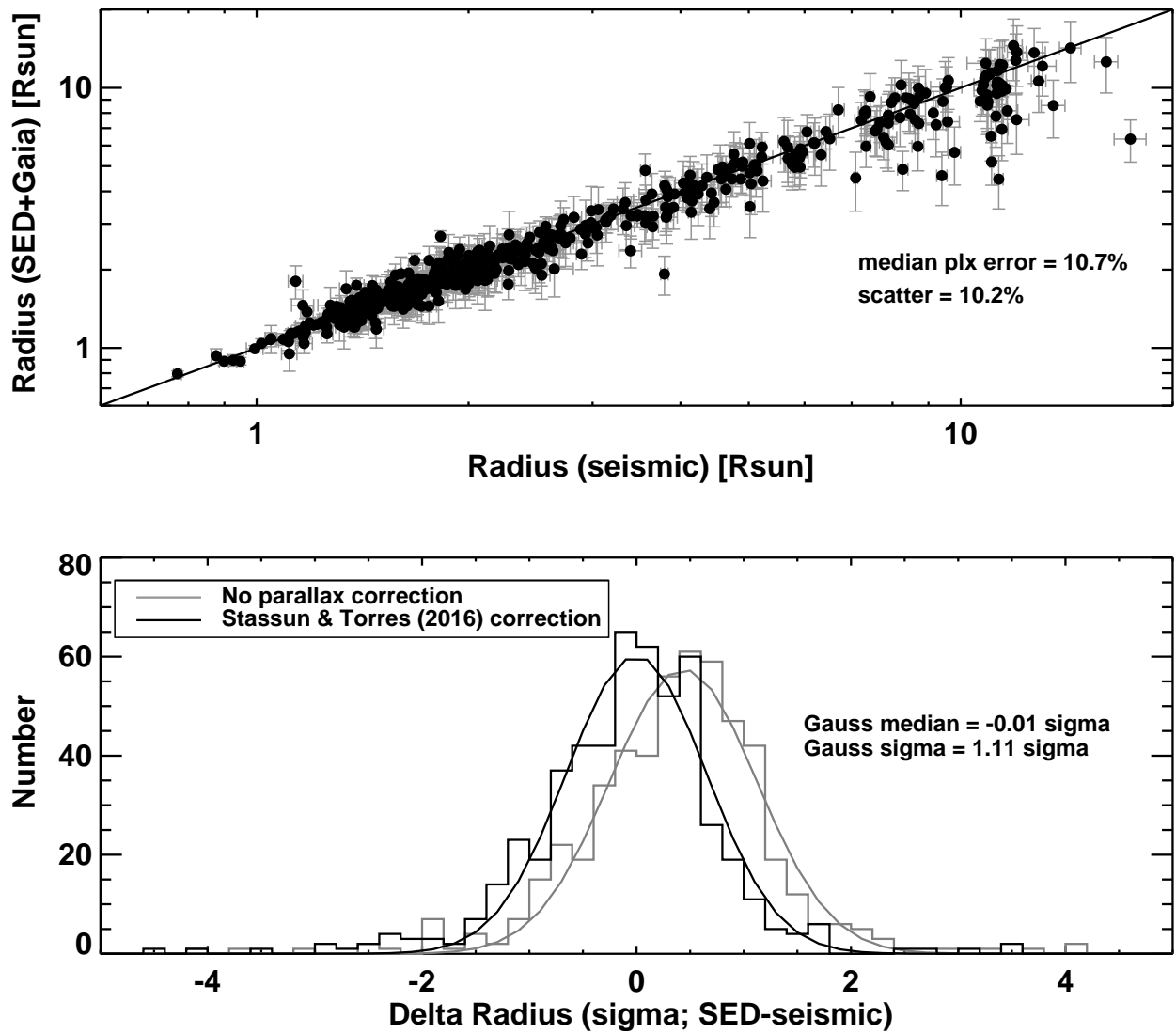}
    \caption{Comparison of stellar radii obtained from SED+parallax versus stellar radii from asteroseismology. (Top:) Direct comparison. (Bottom:) Histogram of differences in units of measurement uncertainty; a small offset is explained by the systematic error in the {\it Gaia\/} DR1 parallaxes reported by \citet{StassunGaiaError:2016}.}
    \label{fig:radii}
\end{figure*}

Figure~\ref{fig:radii} (top) shows that the SED+parallax based \rstar\ agree beautifully with the seismic \rstar, and the scatter of $\sim$10\% is as expected for the typical parallax error in this sample of $\sim$10\%. 

Figure~\ref{fig:radii} (bottom) demonstrates that the residuals between \rstar\ obtained from the two methods are normally distributed as expected. However, there is a small systematic offset apparent. Applying the systematic correction to the {\it Gaia\/} DR1 parallaxes reported by \citet{StassunGaiaError:2016} effectively removes this offset. 
The spread in the residuals is almost exactly that expected for the measurement errors (1.1$\sigma$, where $\sigma$ represents the typical measurement error).

\subsection{Expected Precision of Stellar Masses}\label{subsec:mass}

Finally, we consider the expected precision on \mstar\ that may be achieved through the combination of the granulation-based \logg\ with the SED+parallax based \rstar. Again we utilize the same demonstration sample as above, comparing the \mstar\ inferred from the above results against the asteroseismically determined \mstar. 

\begin{figure*}[!ht]
    \centering
    \includegraphics[trim=75 75 60 85,clip,width=0.85\linewidth]{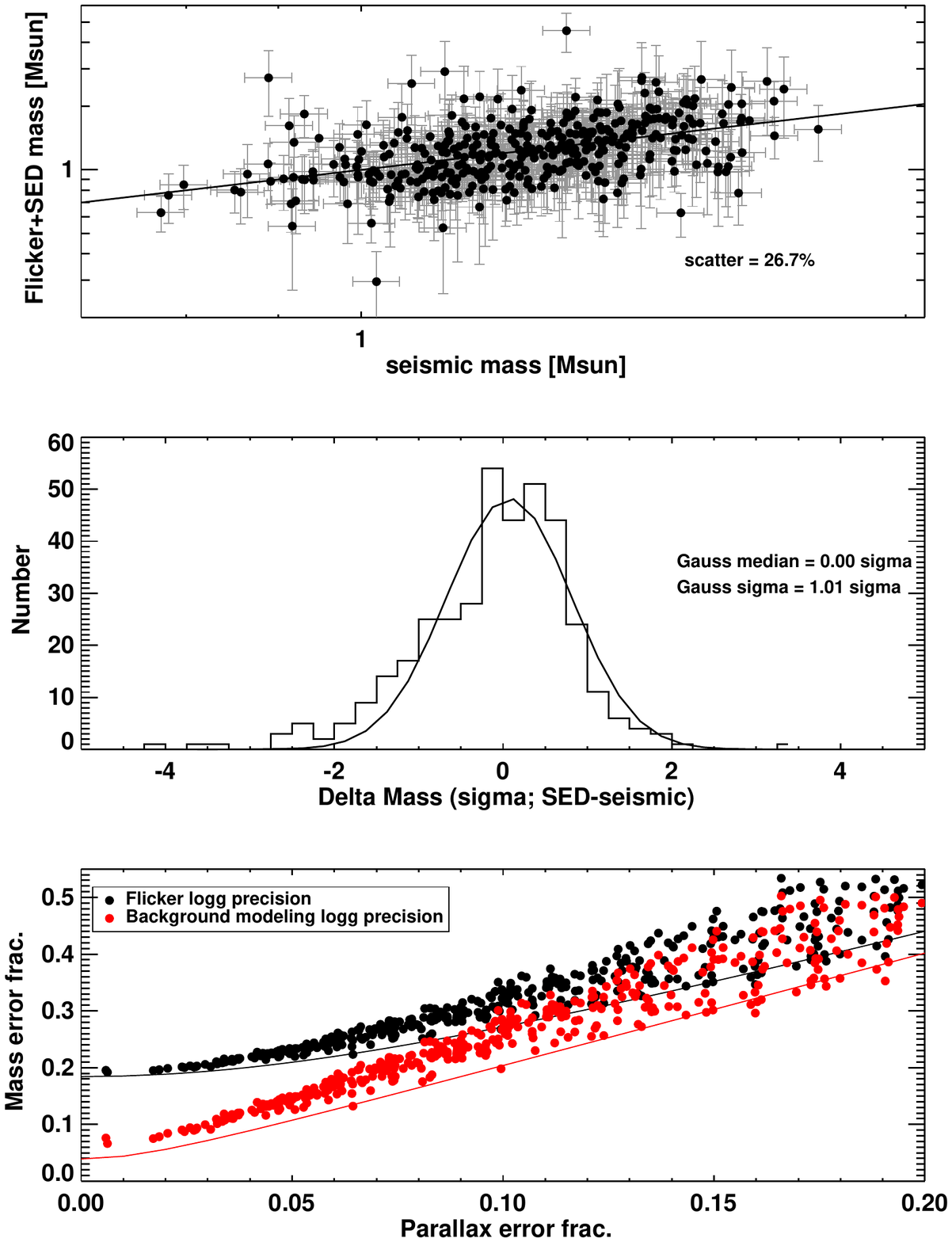}
    \caption{(Top:) Comparison of \mstar\ obtained from \flick-based \logg\ and SED+parallax based \rstar, versus \mstar\ from asteroseismology. (Middle:) Histogram of the residuals from top panel. (Bottom:) Actual \mstar\ precision versus parallax error for \logg\ measured from \flick\ (black) and the same but assuming improved \logg\ precision achievable from granulation background modeling \citep{Corsaro:2017} applied to {\it TESS\/} data (red). Symbols represent actual stars used in this study; solid curves represent expected precision floor based on nominal \logg\ precision (0.08~dex from \flick, 0.02~dex from granulation background).}
    \label{fig:mass}
\end{figure*}

Figure~\ref{fig:mass} (top) shows the direct comparison of \mstar\ from the two methods. The mass estimated from the SED+parallax based \rstar\ (with parallax systematic correction applied) and \flick-based \logg\ compares beautifully with the seismic \mstar. The scatter of $\sim$25\% is as expected for the combination of 0.08~dex \logg\ error from \flick\ and the median parallax error of $\sim$10\% for the sample. 

The \mstar\ residuals are normally distributed (Figure~\ref{fig:mass}, middle), 
and again the spread in the residuals is as expected for the measurement errors.
The \mstar\ uncertainty is dominated by the \flick-based \logg\ error for stars with small parallax errors, and follows the expected error floor (Figure~\ref{fig:mass}, bottom, black). The \mstar\ precision is significantly improved for bright stars if we instead assume the \logg\ precision expected from the granulation background modeling method of \citet{Corsaro:2017}. 
For parallax errors of less than 5\%, as will be the case for most of the {\it TESS\/} stars with {\it Gaia\/} DR2, we can expect \mstar\ errors of less than $\sim$10\%.

\section{Discussion}\label{sec:discussion}

The primary goal of this paper is to explore the upcoming potential of {\it TESS\/} and {\it Gaia}, together with large archival photometric datasets---from the ultraviolet to the mid-infrared---in order to make accurate stellar radius and mass measurements for large numbers of stars---especially single stars---across the sky. 
Indeed, this will enable precise testing of evolutionary models for single stars across the H-R diagram, including the ability to fully characterize and understand the role of magnetic activity on stellar radius inflation, and many other areas of stellar astrophysics that depend on the accuracy of stellar models. 
Stellar models are the main tools for determining the masses and ages of most stars---including the determination of the stellar initial mass function and the star formation history of the galaxy. 

Empirical and accurate determinations of fundamental radii and masses for large numbers of stars across the H-R diagram will inevitably lead to improvements in the stellar models, which rely on empirical measurements of basic stellar properties for calibration.
At the same time, a secondary benefit is to further enable the characterization of extrasolar planets, whose properties depend on knowledge of the host-star properties, which is of course a main objective of the {\it TESS\/} and {\it PLATO\/} missions. 

In this section we discuss the estimated yield of accurate, empirical \rstar\ and \mstar\ via the methods laid out in this work, discuss some example applications of such a large sample of empirical stellar properties, and lastly consider some caveats and limitations of the approach developed here.

\subsection{Estimated yield of stellar radii and masses}

We begin by estimating the number of stars in the {\it TESS\/} Input Catalog (TIC) and in the {\it TESS\/} Candidate Target List (CTL) to which we may 
apply our procedures from \citet{StassunGaiaEB:2016}, \citet{StassunGaiaPlanets:2017}, and \citet{Stevens:2017} in order to obtain \rstar. 

As described in Section~\ref{sec:methods}, this involves measuring \fbol\ and angular radius via the broadband SED, constructed from {\it GALEX}, {\it Gaia}, {\it 2MASS}, and {\it WISE}---spanning a wavelength range 0.15--22~\micron---supplemented with broadband photometric measurements at visible wavelengths from {\it Tycho-2}, {\it APASS}, and/or {\it SDSS}. With the addition of the {\it Gaia\/} DR2 parallax, the angular radius then yields \rstar. 

{For the \mstar\ determination via the granulation-based \logg\ measurement, we require the stars to be cool enough to possess a surface convection zone, i.e., \teff$\lesssim$6750~K. For the \flick-based granulation measurement, we also exclude red giants, given that method's range of applicability \citep[i.e., \logg$\lesssim$3;][]{Bastien:2013,Bastien:2016}. Finally, using the estimated flux contamination of nearby sources as provided by the TIC \citep{StassunTIC:2017}, we select stars whose total estimated flux contamination is less than 10\%, to avoid stars whose SED fitting and/or granulation signals may be compromised by the presence of other signals.}

As shown in Table~\ref{tab:sample}, accurate and empirical measures of \rstar\ should be attainable for nearly 100 million stars possessing {\it Gaia\/} parallaxes and for which SEDs can be constructed from visible to mid-infrared wavelengths. A subset of these, about 28 million, will also have {\it GALEX\/} ultraviolet fluxes which, while helpful especially for hot stars, are not crucial for obtaining reliable \fbol\ for most stars \citep{StassunGaiaEB:2016}.

\begin{center}
	\begin{longtable*}[c]{|c|c|c|c|c|}
 		\hline
 		   & {\it GALEX\/} (UV) & {\it Gaia\/} (Visible) & {\it 2MASS\/} (near-IR) & {\it WISE\/} (mid-IR) \\
 		\hline
 		\rstar\ for TIC stars in {\it Gaia\/} DR-2 	& 28M & 97M		& 448M 	& 311M \\
 		\mstar\ via \flick\ for TIC stars with $T_{\rm mag} < 10.5$ & 16k & 339k & 339k & 332k \\
 		\mstar\ via $b_{\rm meso}$ for CTL stars with $T_{\rm mag} < 7$ & 0.5k & 12k & 12k & 11k \\
 		\mstar\ via $b_{\rm meso}$ for TIC stars with $T_{\rm mag} < 7$ & 1.6k & 34k & 34k & 33k \\
 		\hline
 		\caption{Approximate numbers of stars for which \rstar\ and \mstar\ can be obtained via the methods described in this paper, according to the data available with which to construct SEDs from {\it GALEX}, Visible ({\it Gaia}, SDSS, APASS, Tycho-2), {\it 2MASS}, and {\it WISE}. 
 		\label{tab:sample}}
 	\vspace{-0.25in}
	\end{longtable*}
\end{center}

As shown in Table~\ref{tab:sample}, we estimate that accurate and empirical \mstar\ measurements should be obtainable for {$\sim$300 thousand} {\it TESS\/} stars via \flick-based gravities. These masses should be good to about 25\% (see Section~\ref{sec:results}). In addition, we estimate that a smaller but more accurate and precise set of \mstar\ measurements should be possible via the granulation background modeling method for $\sim$11k bright {\it TESS\/} stars in the CTL 2-min cadence targets, and for another $\sim$33k bright {\it TESS\/} stars in the TIC 30-min cadence targets.

\subsection{Applications of fundamental \mstar\ and \rstar\ measurements with {\it TESS\/} and {\it Gaia}}

\subsubsection{Determination of the relationships between radius inflation, activity, and rotation}

One of the major outstanding puzzles in fundamental stellar physics is the so-called ``radius inflation'' problem---the peculiar trend of some stars of mass $\lesssim 1$\msun\ to have radii that are physically larger by $\sim 5-10$\% relative to the predictions of state-of-the-art stellar models. This phenomenon has been discovered in eclipsing binaries \citep[e.g][]{Lopez-Morales:2007}, statistical studies of open clusters \citep[e.g.][]{Jackson:2016}, on both sides of the fully-convective boundary of $0.35$\msun\ \citep[e.g.][]{Stassun:2012}, and on both the pre-main sequence \citep{Stassun:2014} and main sequence \citep[e.g.][]{Feiden:2012}, demonstrating inflation as a ubiquitous feature of low-mass stellar evolution. 

A precise census of the magnitude of radius inflation as a function of mass, age, and other relevant stellar parameters will be critical for accurate characterization of exoplanet radii, for precise age measurements of young star-forming regions, and for measurements of the stellar initial mass function \citep[e.g.][]{Somers:2015}. Though the term ``inflation'' seems to denote some fault of the stars themselves, the clear implication is missing ingredients in our stellar models. Therefore, unveiling the true mechanism behind radius inflation also promises new revelations about the fundamental physics driving the structure and evolution of stars.

Most radius inflation studies have been carried out with eclipsing binaries, which are rare and costly to analyze. 
The methods outlined in this paper should provide a new avenue for measuring large samples of stellar radii, from which radius inflation measures can be readily derived. 

\begin{figure}[!ht]
\centering
\includegraphics[width=\linewidth,trim=5 2 2 25,clip]{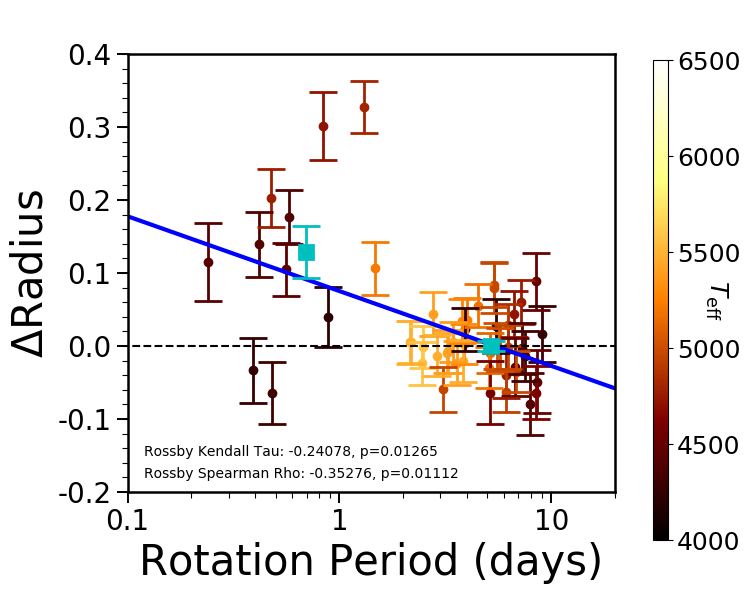}
\caption{Adapted from \citet{Somers:2017}; a comparison between the rotation period of Pleiades stars and their fractional height above appropriately-aged stellar isochrones from \citet{Bressan:2012}. Pleiads rotating slower than 1.5~days show good agreement with predictions, but faster rotating stars are systematically larger by on average 10-20\%. The cyan squares shows the average $\Delta R$ among the slower and faster stars, divided at 1.5~days. The trend is statistically significant according to Kendall's $\tau$ and Spearman's $\rho$ coefficients. This suggests that rapid rotation drives radius inflation, perhaps through the influence of correlated starspots and magnetic activity.
\label{fig:Somers2017}}
\end{figure}

The capacity of this methodology to probe the nature of radius inflation has been demonstrated in \citet{Somers:2017}, who derived empirical radii for dozens of K-type dwarfs in the Pleiades, and determined the magnitude of radius inflation exhibited by each star. They found evidence for a clear connection between rapid rotation ($P_{\rm rot} < 1.5$~d) and significant levels of radius inflation ($\sim$10--20\%), providing some insight into the physical processes at play (see Figure~\ref{fig:Somers2017}). In particular, this preliminary study shows that radius inflation in low-mass stars is connected to rapid stellar rotation---probably because rapid rotation drives a stronger magnetic dynamo---and furthermore provides an empirical calibration of the effect at an age of 120~Myr.

However, the limitations of this sample, namely the small mass range, the solitary age of the cluster, and the low raw numbers, precluded a comprehensive calibration of radius inflation as a function of rotation---this fact has been typical of studies in the field to date.
With the very large number of {\it TESS\/} stars for which \rstar\ and \mstar\ will be measurable (Table~\ref{tab:sample}), this state of affairs is set to radically change, enabling a direct probe of the nature of radius inflation with a sample of unprecedented size and diversity. 
In addition, rotation period measurements for large numbers of {\it TESS\/} stars are available already (Oelkers et al.\ 2017, in preparation) and more will be measurable from the {\it TESS\/} light curves themselves. 
Thus it should become possible to perform comprehensive studies of radius inflation, tracing its magnitude along the mass function and throughout the stellar life cycle.


\subsubsection{Empirical determination of accurate radii and masses of exoplanets}

Accurate, empirical estimates of the radii (\rplanet) and masses (\mplanet) of extrasolar planets are essential for a broad range of exoplanet science. These parameters yield the bulk density of an exoplanet, and thus broadly categorize its nature (gas giant, ice giant, mini-Neptune, rocky planet, etc). 
Planet masses and radii can also provide important insight into both the physics of planetary atmospheres and interiors, and the physics of planet formation and evolution. 
For example, estimates of the masses and radii of low-mass planets ($M_p\la 10 M_\oplus$) detected via {\it Kepler\/} have uncovered an apparent dichotomy in the properties of planets with radii $\la 1.5~R_\oplus$ compared to those larger than this \citep{rogers2015}, such that larger planets appear to have significant hydrogen and helium envelopes whereas  smaller planets appear to be much more similar to the terrestrial planets in our solar system.  

As is well known, in order to reliably estimate \rplanet\ and \mplanet, one must have an accurate measure of \rstar\ and \mstar.
Up until now, these observables of the host stars have rarely been obtained empirically.  
Instead, most studies have used theoretical models and/or empirically-calibrated relations between other observable properties of the star (e.g., main-sequence \rstar--\teff\ relations). Stellar evolution models and empirical relations are reasonably well understood, 
nevertheless the models are subject to uncertainties in input physics and in second-order parameters (e.g., stellar rotation), and empirical relations are subject to calibration uncertainties. Such estimates of stellar parameters, while precise, are therefore not necessarily accurate. One demonstration of this is KELT-6b \citep{Collins:2014}, where the parameters inferred using the Yonsei-Yale model isochrones disagreed by as much as $4\sigma$ relative to the \citet{Torres:2010} empirical relations, likely due to the fact that neither the isochrones nor the empirical relations are well-calibrated at low metallicities. 

\begin{figure}[!ht]
    \centering
    \includegraphics[width=\linewidth,trim=10 10 10 50,clip]{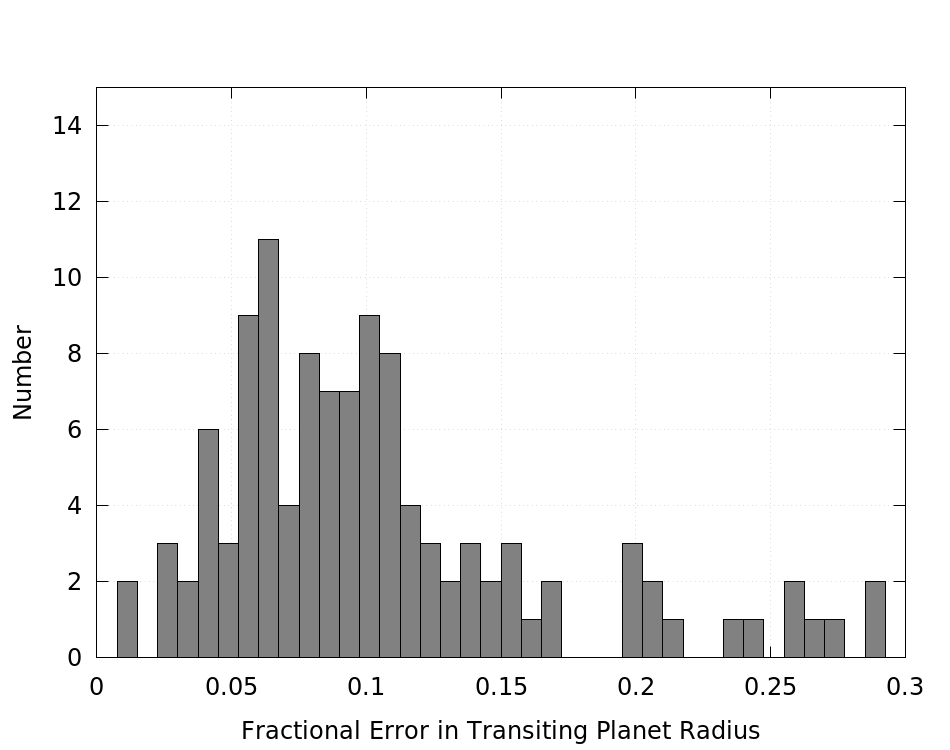}
    \includegraphics[width=\linewidth,trim=10 10 10 50,clip]{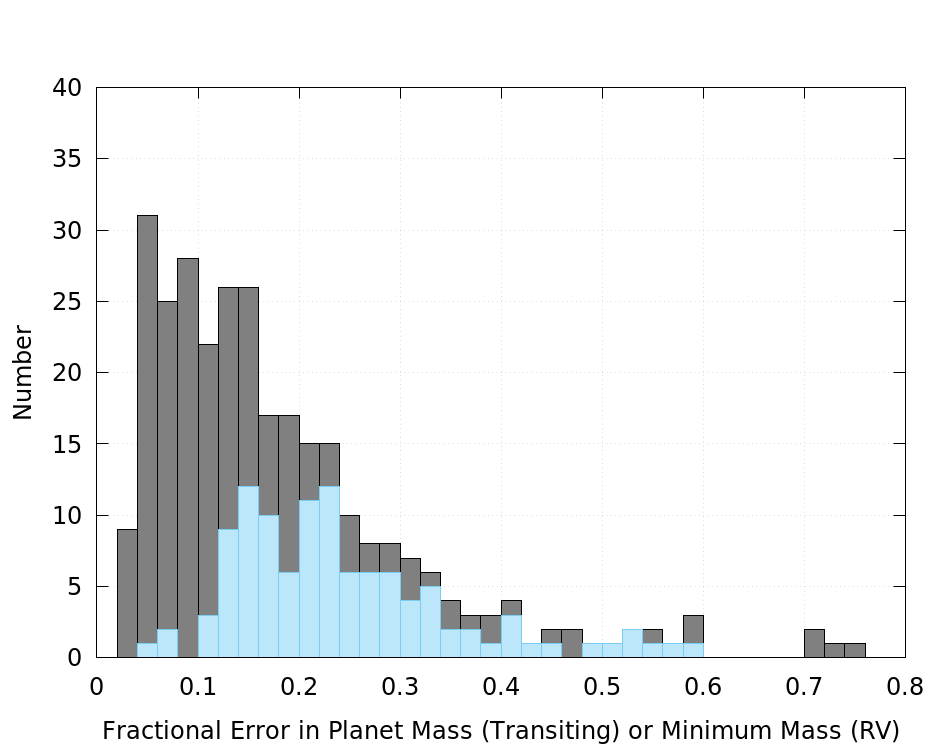}
    \caption{Distributions of fractional uncertainties on \rplanet\ (top) and \mplanet\ (\mplanet\sini\ for the RV planets) (bottom) determined from the empirical \rstar\ and other direct observables. Transiting planets are represented in blue in the right panel. Reproduced from \citet{StassunGaiaPlanets:2017}.}
    \label{fig:trans_planets}
\end{figure}

In \citet{StassunGaiaPlanets:2017} we developed a methodology that combines empirical measurements of \rstar---obtained using the method described in Section~\ref{sec:methods}---with empirical observables of transiting exoplanets (such as the transit depth, \depth) to empirically determine \rplanet\ and \mplanet\ (see Fig.~\ref{fig:trans_planets}). 
The \citet{StassunGaiaPlanets:2017} analysis used only direct, empirical observables and included an empirically calibrated covariance matrix for properly and accurately propagating uncertainties. 

In particular, for transiting planets we determine the stellar density, \rhostar, from the transit model parameter \ar\ and the orbital period, $P$, through the relation
$\rhostar = \frac{3\pi}{GP^2}(\ar)^3$.
Combining \rhostar\ with the empirically determined \rstar\ provides a direct measure of \mstar\ akin to that obtained via \logg\ as described in Section~\ref{sec:methods}.
From the empirically calculated \rstar, \rplanet\ follows directly via $\rprstar=\sqrt{\depth}$. Similarly, from the empirically calculated \mstar, \mplanet\ follows directly via
$\mplanet = \frac{\rvamp\sqrt{1-e^2}}{\sini}\left(\frac{P}{2\pi G}\right)^{1/3}\mstar^{2/3}$,
where $e$ is the orbital eccentricity and \rvamp\ is the orbital RV semi-amplitude.
Of course, it is also possible to empirically measure \mplanet\ for non-transiting (i.e., radial-velocity) planets by again using the empirical \rstar\ together with the granulation-based \logg\ to measure \mstar\ (see Fig.~\ref{fig:trans_planets}b). 

\citet{StassunGaiaPlanets:2017} achieved a typical accuracy of $\sim$10\% in \rplanet\ and $\sim$20\% in \mplanet, limited by the {\it Gaia\/} DR1 parallaxes then available (see Fig.~\ref{fig:fbol_vs_chi2}); with the significantly improved parallaxes expected from {\it Gaia\/} DR2, the stellar and planet radii and masses should achieve an accuracy of $\approx$3\% and $\approx$5\%, respectively \citep{StassunGaiaPlanets:2017}.

\subsection{Potential Sources of Systematic Uncertainty and Mitigation Strategy}

The methodologies outlined in this paper to determine \rstar\ and \mstar\ for large numbers of stars from {\it TESS\/} and {\it Gaia\/} are relatively straightforward, and as we have described, essentially empirical.  Nevertheless, as with nearly all measurements made in astronomy or any other scientific discipline, they cannot in truth be described as purely empirical.  Rather, we must make some simplifying assumptions and rely on some theory, models, and extrapolation, at least to some degree. 

Here we discuss some of the potential sources of systematic uncertainty stemming from our methodology that may affect the final achievable accuracy.  We also outline ways in which these can be checked and mitigated, using data available now and in the future.  

\subsubsection{Bolometric Flux} 

The first step in our analysis is to estimate the de-extincted stellar \fbol.  As discussed above, this is done by assembling archival broadband fluxes from a number of sources over a wavelength range of (at most) 0.15--22~\micron.  We then fit SEDs derived from stellar atmosphere model to these fluxes, with $\fbol$ and $A_V$, and, if no spectroscopic estimate is available, $\teff$, as free parameters.  There are a number of sources of uncertainties that can be introduced when estimating $\fbol$ in this way. 

First, the theoretical SEDs formally depend on $\loggstar$ and [Fe/H] as well.  However, the shape of the SEDs are generally weak functions of these parameters, at least over the wavelengths where the majority of the flux is emitted and for typical ranges of these parameters.  Nevertheless, one can estimate the magnitude of the error introduced by assuming fiducial values of $\loggstar$ and [Fe/H] using stars for which these parameters have independent measurements (e.g., via high resolution spectra). \citet{StassunGaiaEB:2016} found the net effect on \fbol\ to be of order 1\% for the vast majority of stellar \teff\ and [Fe/H] encountered in the Milky Way. 

Second, the reliance on stellar atmospheres to effectively interpolate and extrapolate between and beyond the broadband flux measurements means that the estimate of $\fbol$ is not entirely model-independent.  However, we have tested the effect of using two different model atmospheres \citep{Kurucz2013,Baraffe1998} for a number of typical cases, and found the difference in the estimated $\fbol$ from the two models to be below the typical statistical uncertainty \citep{Stevens:2017}. 
In the future, {\it Gaia\/} spectrophotometry will enable a more direct measurement of $\fbol$ in the 0.3--1~\micron\ range.

Third, the effect of extinction must be accounted for in order to estimate the true $\fbol$. This requires adopting a parameterized extinction law (e.g., \citealt{Cardelli:1989}), a value for the ratio of total-to-selective extinction $R_V$, and fitting for the $V$-band extinction $A_V$.  For most stars to be observed by {\it TESS}, leverage on the extinction primarily comes from comparing the long wavelength {\it WISE\/} fluxes, which are essentially unextincted for the majority of the stars of interest, to the broadband optical fluxes.  While we do not expect the extinction law nor $R_V$ to deviate significantly from the standard Cardelli law or $R_V=3.1$, \citet{StassunGaiaEB:2016} test the degree to which estimates of $\fbol$ change with different assumptions about the form of the extinction law, again finding the effect to be on the order of at most a few percent for the full range of $R_V$ expected in the Milky Way.  We note that, should it be selected, the SPHEREx mission \citep{Dore2014} will provide low-resolution spectrophotometry between 1--5~\micron\ which, when combined with {\it Gaia\/} spectrophotometry, will enable a {\it direct} measurement of $\ga 90-95\%$ of the flux of late F, G, and K stars, and, combined with stellar atmosphere models, a simultaneous estimate of $\fbol$ and the extinction as a function of wavelength, without requiring a prior assumption about the form of the extinction law. 

\subsubsection{Effective Temperature} 

Formally, $\teff$ is a defined quantity: 
$\teff\equiv \left({L_{\rm bol}}/{4\pi\sigma_{\rm SB}\rstar^2}\right)^{1/4}$.
However, in our methodology we use $\teff$ as an {\it input} to determine \rstar.  

Measurements of $\teff$ from high-resolution stellar spectra typically rely on stellar atmosphere models, which are normalized such that the above identity holds.  The most sophisticated of these models do not assume plane-parallel atmospheres, and thus account for the effect of limb darkening on the stellar spectra as well.  Nevertheless, the choice of the spectral lines used to estimate $\teff$ can affect its inferred value, since different spectral lines (and, indeed, different parts of the lines) originate from different depths in the stellar photosphere.  It is general practice to use those lines that yield values of $\teff$ that best reproduce the definition above for the model adopted, as calibrated using standard stars with accurate and precise angular diameter measurements (see, e.g., Table \ref{tab:boyajian}).  

There is not much in practice that can be done to measure \teff\ with fundamental accuracy, or indeed to avoid the definitional nature of \teff\ as a quantity. Comparisons of spectroscopic \teff\ obtained by various spectroscopic methods as well as from independent methods such as colors, generally find systematic differences in \teff\ scales on the order of 100~K \citep[e.g.,][]{Huber:2017}. This is a $\sim$2\% effect for cool stars and $\sim$1\% for hot stars, which may fundamentally limit the accuracy of \rstar\ determinations to a few percent for most stars.

\subsubsection{Distance}  

The {\it Gaia\/} parallaxes are an essential ingredient in our methodology to determine \rstar\ and then \mstar. Here the critical assumption is that the {\it Gaia\/} parallaxes themselves do not contain significant systematic uncertainties as compared to the quoted statistical precisions. Of course, it is well known that there can be many sources of systematic uncertainty when measuring parallaxes: unrecognized binary companions, Lucy-Sweeney bias \citep{Lucy1971}, Lutz-Kelker bias \citep{Lutz1973}, and potential systematic errors in the {\it Gaia\/} data reduction methodology itself.  

Fortunately, there are methods for independently assessing the accuracy of the trigonometric parallaxes. For example, \citet{StassunGaiaError:2016} found a systematic offset of $-$0.25~mas in the {\it Gaia\/} DR1 parallaxes ({\it Gaia\/} parallaxes slightly too small) by comparison to distances inferred for a set of benchmark double-lined eclipsing binaries \citep[see also, e.g.,][]{Davies:2017}.

\section{Summary and Conclusions}\label{sec:summary}
In this paper, we have sought to lay out a methodology by which radii (\rstar) and masses (\mstar) of stars may be determined empirically and accurately with the data that will soon become available for millions of stars across the sky from {\it TESS\/} and {\it Gaia}. Importantly, as it does not rely upon the presence of an orbiting, eclipsing, or transiting body, the methodology provides a path to \rstar\ and \mstar\ determinations for {\it single\/} stars. 

In brief, the method involves the determination of \rstar\ from the bolometric flux at Earth (\fbol) obtained via the broadband SED, the stellar \teff\ obtained spectroscopically or else also from the SED, and the parallax; the determination of the stellar surface gravity (\logg) from the granulation-driven brightness variations in the light curve; and then \mstar\ from the combination of \rstar\ and \logg. 

Using a sample of \nsamplefin\ stars in the {\it Kepler\/} field for which the above measures are available as well asteroseismic gold-standard \rstar\ and \mstar\ determinations for comparison, we find that the method faithfully reproduces \rstar\ and \mstar, good to $\approx$10\% and $\approx$25\%, respectively. The accuracy on \rstar\ is at present limited by the precision of the {\it Gaia\/} DR1 parallaxes, and the accuracy on \mstar\ is at present limited by the precision of granulation ``flicker" based \logg. We show that with improvements in the parallaxes expected from {\it Gaia\/} DR2, and with improvements in the granulation-based \logg\ via Fourier background modeling techniques \citep[e.g.,][]{Corsaro:2017} as applied to {\it TESS}, the accuracy of the \rstar\ and \mstar\ determinations can be improved to $\approx$3\% and $\approx$10\%, respectively. 

From the {\it TESS\/} Input Catalog \citep{StassunTIC:2017} we estimate that this methodology may be applied to as many as $\sim$100 million {\it TESS\/} stars for determination of accurate and empirical \rstar, and to as many as {$\sim$300 thousand} {\it TESS\/} stars for determination of accurate and empirical \mstar.

\acknowledgments
We thank R.~Oelkers for assistance with the {\it TESS\/} Input Catalog. {We are grateful to D.\ Huber and the anonymous referee for helpful criticisms that improved the paper.} This work has made use of the Filtergraph data visualization service \citep{Burger2013}, developed through support from the Vanderbilt Initiative in Data-intensive Astrophysics (VIDA) and the Vanderbilt Center for Autism \& Innovation. K.G.S.\ acknowledges support from NSF PAARE grant AST-1358862. 
E.C.\ is funded by the European Union's Horizon 2020 research and innovation program under the Marie Sklodowska-Curie grant agreement No.~664931.

\bibliography{main}

\begin{thebibliography}{}
\expandafter\ifx\csname natexlab\endcsname\relax\def\natexlab#1{#1}\fi
\providecommand{\url}[1]{\href{#1}{#1}}

\bibitem[{{Bailer-Jones}(2015)}]{BailerJones:2015}
{Bailer-Jones}, C.~A.~L. 2015, \pasp, 127, 994

\bibitem[{{Baraffe} {et~al.}(1998){Baraffe}, {Chabrier}, {Allard}, \&
  {Hauschildt}}]{Baraffe1998}
{Baraffe}, I., {Chabrier}, G., {Allard}, F., \& {Hauschildt}, P.~H. 1998, \aap,
  337, 403

\bibitem[{{Bastien} {et~al.}(2013){Bastien}, {Stassun}, {Basri}, \&
  {Pepper}}]{Bastien:2013}
{Bastien}, F.~A., {Stassun}, K.~G., {Basri}, G., \& {Pepper}, J. 2013, \nat,
  500, 427

\bibitem[{{Bastien} {et~al.}(2016){Bastien}, {Stassun}, {Basri}, \&
  {Pepper}}]{Bastien:2016}
---. 2016, \apj, 818, 43

\bibitem[{{Birkby} {et~al.}(2012){Birkby}, {Nefs}, {Hodgkin}, {Kov{\'a}cs},
  {Sip{\H o}cz}, {Pinfield}, {Snellen}, {Mislis}, {Murgas}, {Lodieu}, {de
  Mooij}, {Goulding}, {Cruz}, {Stoev}, {Cappetta}, {Palle}, {Barrado},
  {Saglia}, {Martin}, \& {Pavlenko}}]{Birkby2012}
{Birkby}, J., {Nefs}, B., {Hodgkin}, S., {et~al.} 2012, \mnras, 426, 1507

\bibitem[{{Boyajian} {et~al.}(2015){Boyajian}, {von Braun}, {Feiden}, {Huber},
  {Basu}, {Demarque}, {Fischer}, {Schaefer}, {Mann}, {White}, {Maestro},
  {Brewer}, {Lamell}, {Spada}, {L{\'o}pez-Morales}, {Ireland}, {Farrington},
  {van Belle}, {Kane}, {Jones}, {ten Brummelaar}, {Ciardi}, {McAlister},
  {Ridgway}, {Goldfinger}, {Turner}, \& {Sturmann}}]{Boyajian:2015}
{Boyajian}, T., {von Braun}, K., {Feiden}, G.~A., {et~al.} 2015, \mnras, 447,
  846

\bibitem[{{Boyajian} {et~al.}(2012{\natexlab{a}}){Boyajian}, {McAlister}, {van
  Belle}, {Gies}, {ten Brummelaar}, {von Braun}, {Farrington}, {Goldfinger},
  {O'Brien}, {Parks}, {Richardson}, {Ridgway}, {Schaefer}, {Sturmann},
  {Sturmann}, {Touhami}, {Turner}, \& {White}}]{Boyajian2012a}
{Boyajian}, T.~S., {McAlister}, H.~A., {van Belle}, G., {et~al.}
  2012{\natexlab{a}}, \apj, 746, 101

\bibitem[{{Boyajian} {et~al.}(2012{\natexlab{b}}){Boyajian}, {von Braun}, {van
  Belle}, {McAlister}, {ten Brummelaar}, {Kane}, {Muirhead}, {Jones}, {White},
  {Schaefer}, {Ciardi}, {Henry}, {L{\'o}pez-Morales}, {Ridgway}, {Gies}, {Jao},
  {Rojas-Ayala}, {Parks}, {Sturmann}, {Sturmann}, {Turner}, {Farrington},
  {Goldfinger}, \& {Berger}}]{Boyajian2012b}
{Boyajian}, T.~S., {von Braun}, K., {van Belle}, G., {et~al.}
  2012{\natexlab{b}}, \apj, 757, 112

\bibitem[{{Bressan} {et~al.}(2012){Bressan}, {Marigo}, {Girardi}, {Salasnich},
  {Dal Cero}, {Rubele}, \& {Nanni}}]{Bressan:2012}
{Bressan}, A., {Marigo}, P., {Girardi}, L., {et~al.} 2012, \mnras, 427, 127

\bibitem[{{Brown} {et~al.}(1991){Brown}, {Gilliland}, {Noyes}, \&
  {Ramsey}}]{Brown91}
{Brown}, T.~M., {Gilliland}, R.~L., {Noyes}, R.~W., \& {Ramsey}, L.~W. 1991,
  \apj, 368, 599

\bibitem[{{Burger} {et~al.}(2013){Burger}, {Stassun}, {Pepper}, {Siverd},
  {Paegert}, {De Lee}, \& {Robinson}}]{Burger2013}
{Burger}, D., {Stassun}, K.~G., {Pepper}, J., {et~al.} 2013, Astronomy and
  Computing, 2, 40

\bibitem[{{Campante} {et~al.}(2016){Campante}, {Schofield}, {Kuszlewicz},
  {Bouma}, {Chaplin}, {Huber}, {Christensen-Dalsgaard}, {Kjeldsen}, {Bossini},
  {North}, {Appourchaux}, {Latham}, {Pepper}, {Ricker}, {Stassun},
  {Vanderspek}, \& {Winn}}]{Campante:2016}
{Campante}, T.~L., {Schofield}, M., {Kuszlewicz}, J.~S., {et~al.} 2016, \apj,
  830, 138

\bibitem[{{Cardelli} {et~al.}(1989){Cardelli}, {Clayton}, \&
  {Mathis}}]{Cardelli:1989}
{Cardelli}, J.~A., {Clayton}, G.~C., \& {Mathis}, J.~S. 1989, \apj, 345, 245

\bibitem[{{Collins} {et~al.}(2014){Collins}, {Eastman}, {Beatty}, {Siverd},
  {Gaudi}, {Pepper}, {Kielkopf}, {Johnson}, {Howard}, {Fischer}, {Manner},
  {Bieryla}, {Latham}, {Fulton}, {Gregorio}, {Buchhave}, {Jensen}, {Stassun},
  {Penev}, {Crepp}, {Hinkley}, {Street}, {Cargile}, {Mack}, {Oberst}, {Avril},
  {Mellon}, {McLeod}, {Penny}, {Stefanik}, {Berlind}, {Calkins}, {Mao},
  {Richert}, {DePoy}, {Esquerdo}, {Gould}, {Marshall}, {Oelkers}, {Pogge},
  {Trueblood}, \& {Trueblood}}]{Collins:2014}
{Collins}, K.~A., {Eastman}, J.~D., {Beatty}, T.~G., {et~al.} 2014, \aj, 147,
  39

\bibitem[{{Corsaro} \& {De Ridder}(2014)}]{Corsaro14}
{Corsaro}, E., \& {De Ridder}, J. 2014, \aap, 571, A71

\bibitem[{{Corsaro} {et~al.}(2015){Corsaro}, {De Ridder}, \&
  {Garc{\'{\i}}a}}]{Corsaro15}
{Corsaro}, E., {De Ridder}, J., \& {Garc{\'{\i}}a}, R.~A. 2015, \aap, 579, A83

\bibitem[{{Corsaro} {et~al.}(2017){Corsaro}, {Mathur}, {Garc{\'{\i}}a},
  {Gaulme}, {Pinsonneault}, {Stassun}, {Stello}, {Tayar}, {Trampedach},
  {Jiang}, {Nitschelm}, \& {Salabert}}]{Corsaro:2017}
{Corsaro}, E., {Mathur}, S., {Garc{\'{\i}}a}, R.~A., {et~al.} 2017, \aap, 605,
  A3

\bibitem[{{Davies} {et~al.}(2017){Davies}, {Lund}, {Miglio}, {Elsworth},
  {Kuszlewicz}, {North}, {Rendle}, {Chaplin}, {Rodrigues}, {Campante},
  {Girardi}, {Hale}, {Hall}, {Jones}, {Kawaler}, {Roxburgh}, \&
  {Schofield}}]{Davies:2017}
{Davies}, G.~R., {Lund}, M.~N., {Miglio}, A., {et~al.} 2017, \aap, 598, L4

\bibitem[{{Dor{\'e}} {et~al.}(2014){Dor{\'e}}, {Bock}, {Ashby}, {Capak},
  {Cooray}, {de Putter}, {Eifler}, {Flagey}, {Gong}, {Habib}, {Heitmann},
  {Hirata}, {Jeong}, {Katti}, {Korngut}, {Krause}, {Lee}, {Masters},
  {Mauskopf}, {Melnick}, {Mennesson}, {Nguyen}, {{\"O}berg}, {Pullen},
  {Raccanelli}, {Smith}, {Song}, {Tolls}, {Unwin}, {Venumadhav}, {Viero},
  {Werner}, \& {Zemcov}}]{Dore2014}
{Dor{\'e}}, O., {Bock}, J., {Ashby}, M., {et~al.} 2014, ArXiv e-prints,
  arXiv:1412.4872

\bibitem[{{Feiden} \& {Chaboyer}(2012)}]{Feiden:2012}
{Feiden}, G.~A., \& {Chaboyer}, B. 2012, \apj, 757, 42

\bibitem[{{Gaia Collaboration} {et~al.}(2016){Gaia Collaboration}, {Brown},
  {Vallenari}, {Prusti}, {de Bruijne}, {Mignard}, {Drimmel}, \&
  {co-authors}}]{GaiaDR1}
{Gaia Collaboration}, {Brown}, A.~G.~A., {Vallenari}, A., {et~al.} 2016, ArXiv
  e-prints, arXiv:1609.04172

\bibitem[{{Handberg} \& {Campante}(2011)}]{Handberg:2011}
{Handberg}, R., \& {Campante}, T.~L. 2011, \aap, 527, A56

\bibitem[{{Harvey}(1985)}]{Harvey85}
{Harvey}, J. 1985, in ESA Special Publication, Vol. 235, Future Missions in
  Solar, Heliospheric \& Space Plasma Physics, ed. E.~{Rolfe} \& B.~{Battrick}

\bibitem[{{Huber} {et~al.}(2017){Huber}, {Zinn}, {Bojsen-Hansen},
  {Pinsonneault}, {Sahlholdt}, {Serenelli}, {Silva Aguirre}, {Stassun},
  {Stello}, {Tayar}, {Bastien}, {Bedding}, {Buchhave}, {Chaplin}, {Davies},
  {Garc{\'{\i}}a}, {Latham}, {Mathur}, {Mosser}, \& {Sharma}}]{Huber:2017}
{Huber}, D., {Zinn}, J., {Bojsen-Hansen}, M., {et~al.} 2017, \apj, 844, 102

\bibitem[{{Jackson} {et~al.}(2016){Jackson}, {Jeffries}, {Randich},
  {Bragaglia}, {Carraro}, {Costado}, {Flaccomio}, {Lanzafame}, {Lardo},
  {Monaco}, {Morbidelli}, {Smiljanic}, \& {Zaggia}}]{Jackson:2016}
{Jackson}, R.~J., {Jeffries}, R.~D., {Randich}, S., {et~al.} 2016, \aap, 586,
  A52

\bibitem[{{Kallinger} {et~al.}(2016){Kallinger}, {Hekker}, {Garcia}, {Huber},
  \& {Matthews}}]{Kallinger16}
{Kallinger}, T., {Hekker}, S., {Garcia}, R.~A., {Huber}, D., \& {Matthews},
  J.~M. 2016, Science Advances, 2, 1500654

\bibitem[{{Kallinger} {et~al.}(2014){Kallinger}, {De Ridder}, {Hekker},
  {Mathur}, {Mosser}, {Gruberbauer}, {Garc{\'{\i}}a}, {Karoff}, \&
  {Ballot}}]{Kallinger14}
{Kallinger}, T., {De Ridder}, J., {Hekker}, S., {et~al.} 2014, \aap, 570, A41

\bibitem[{{Kurucz}(2013)}]{Kurucz2013}
{Kurucz}, R.~L. 2013, {ATLAS12: Opacity sampling model atmosphere program},
  Astrophysics Source Code Library, , , ascl:1303.024

\bibitem[{{L{\'o}pez-Morales}(2007)}]{Lopez-Morales:2007}
{L{\'o}pez-Morales}, M. 2007, \apj, 660, 732

\bibitem[{{Lucy} \& {Sweeney}(1971)}]{Lucy1971}
{Lucy}, L.~B., \& {Sweeney}, M.~A. 1971, \aj, 76, 544

\bibitem[{{Lutz} \& {Kelker}(1973)}]{Lutz1973}
{Lutz}, T.~E., \& {Kelker}, D.~H. 1973, \pasp, 85, 573

\bibitem[{{Mann} {et~al.}(2015){Mann}, {Feiden}, {Gaidos}, {Boyajian}, \& {von
  Braun}}]{Mann2015}
{Mann}, A.~W., {Feiden}, G.~A., {Gaidos}, E., {Boyajian}, T., \& {von Braun},
  K. 2015, \apj, 804, 64

\bibitem[{{Mathur} {et~al.}(2011){Mathur}, {Hekker}, {Trampedach}, {Ballot},
  {Kallinger}, {Buzasi}, {Garc{\'{\i}}a}, {Huber}, {Jim{\'e}nez}, {Mosser},
  {Bedding}, {Elsworth}, {R{\'e}gulo}, {Stello}, {Chaplin}, {De Ridder},
  {Hale}, {Kinemuchi}, {Kjeldsen}, {Mullally}, \& {Thompson}}]{Mathur11}
{Mathur}, S., {Hekker}, S., {Trampedach}, R., {et~al.} 2011, \apj, 741, 119

\bibitem[{{Morales} {et~al.}(2008){Morales}, {Ribas}, \&
  {Jordi}}]{Morales:2008}
{Morales}, J.~C., {Ribas}, I., \& {Jordi}, C. 2008, \aap, 478, 507

\bibitem[{{Privitera} {et~al.}(2016){Privitera}, {Meynet}, {Eggenberger},
  {Vidotto}, {Villaver}, \& {Bianda}}]{Privitera:2016}
{Privitera}, G., {Meynet}, G., {Eggenberger}, P., {et~al.} 2016, \aap, 591, A45

\bibitem[{{Rauer} {et~al.}(2014){Rauer}, {Catala}, {Aerts}, {Appourchaux},
  {Benz}, {Brandeker}, {Christensen-Dalsgaard}, {Deleuil}, {Gizon}, {Goupil},
  {G{\"u}del}, {Janot-Pacheco}, {Mas-Hesse}, {Pagano}, {Piotto}, {Pollacco},
  {Santos}, {Smith}, {Su{\'a}rez}, {Szab{\'o}}, {Udry}, {Adibekyan}, {Alibert},
  {Almenara}, {Amaro-Seoane}, {Eiff}, {Asplund}, {Antonello}, {Barnes},
  {Baudin}, {Belkacem}, {Bergemann}, {Bihain}, {Birch}, {Bonfils}, {Boisse},
  {Bonomo}, {Borsa}, {Brand{\~a}o}, {Brocato}, {Brun}, {Burleigh}, {Burston},
  {Cabrera}, {Cassisi}, {Chaplin}, {Charpinet}, {Chiappini}, {Church},
  {Csizmadia}, {Cunha}, {Damasso}, {Davies}, {Deeg}, {D{\'{\i}}az}, {Dreizler},
  {Dreyer}, {Eggenberger}, {Ehrenreich}, {Eigm{\"u}ller}, {Erikson}, {Farmer},
  {Feltzing}, {de Oliveira Fialho}, {Figueira}, {Forveille}, {Fridlund},
  {Garc{\'{\i}}a}, {Giommi}, {Giuffrida}, {Godolt}, {Gomes da Silva},
  {Granzer}, {Grenfell}, {Grotsch-Noels}, {G{\"u}nther}, {Haswell}, {Hatzes},
  {H{\'e}brard}, {Hekker}, {Helled}, {Heng}, {Jenkins}, {Johansen},
  {Khodachenko}, {Kislyakova}, {Kley}, {Kolb}, {Krivova}, {Kupka}, {Lammer},
  {Lanza}, {Lebreton}, {Magrin}, {Marcos-Arenal}, {Marrese}, {Marques},
  {Martins}, {Mathis}, {Mathur}, {Messina}, {Miglio}, {Montalban}, {Montalto},
  {Monteiro}, {Moradi}, {Moravveji}, {Mordasini}, {Morel}, {Mortier},
  {Nascimbeni}, {Nelson}, {Nielsen}, {Noack}, {Norton}, {Ofir}, {Oshagh},
  {Ouazzani}, {P{\'a}pics}, {Parro}, {Petit}, {Plez}, {Poretti}, {Quirrenbach},
  {Ragazzoni}, {Raimondo}, {Rainer}, {Reese}, {Redmer}, {Reffert},
  {Rojas-Ayala}, {Roxburgh}, {Salmon}, {Santerne}, {Schneider}, {Schou},
  {Schuh}, {Schunker}, {Silva-Valio}, {Silvotti}, {Skillen}, {Snellen}, {Sohl},
  {Sousa}, {Sozzetti}, {Stello}, {Strassmeier}, {{\v S}vanda}, {Szab{\'o}},
  {Tkachenko}, {Valencia}, {Van Grootel}, {Vauclair}, {Ventura}, {Wagner},
  {Walton}, {Weingrill}, {Werner}, {Wheatley}, \& {Zwintz}}]{Rauer:2014}
{Rauer}, H., {Catala}, C., {Aerts}, C., {et~al.} 2014, Experimental Astronomy,
  38, 249

\bibitem[{{Ricker} {et~al.}(2015){Ricker}, {Winn}, {Vanderspek}, {Latham},
  {Bakos}, {Bean}, {Berta-Thompson}, {Brown}, {Buchhave}, {Butler}, {Butler},
  {Chaplin}, {Charbonneau}, {Christensen-Dalsgaard}, {Clampin}, {Deming},
  {Doty}, {De Lee}, {Dressing}, {Dunham}, {Endl}, {Fressin}, {Ge}, {Henning},
  {Holman}, {Howard}, {Ida}, {Jenkins}, {Jernigan}, {Johnson}, {Kaltenegger},
  {Kawai}, {Kjeldsen}, {Laughlin}, {Levine}, {Lin}, {Lissauer}, {MacQueen},
  {Marcy}, {McCullough}, {Morton}, {Narita}, {Paegert}, {Palle}, {Pepe},
  {Pepper}, {Quirrenbach}, {Rinehart}, {Sasselov}, {Sato}, {Seager},
  {Sozzetti}, {Stassun}, {Sullivan}, {Szentgyorgyi}, {Torres}, {Udry}, \&
  {Villasenor}}]{Ricker:2015}
{Ricker}, G.~R., {Winn}, J.~N., {Vanderspek}, R., {et~al.} 2015, Journal of
  Astronomical Telescopes, Instruments, and Systems, 1, 014003

\bibitem[{{Rogers}(2015)}]{rogers2015}
{Rogers}, L.~A. 2015, \apj, 801, 41

\bibitem[{{Somers} \& {Pinsonneault}(2015)}]{Somers:2015}
{Somers}, G., \& {Pinsonneault}, M.~H. 2015, \apj, 807, 174

\bibitem[{{Somers} \& {Stassun}(2017)}]{Somers:2017}
{Somers}, G., \& {Stassun}, K.~G. 2017, \aj, 153, 101

\bibitem[{{Soubiran} {et~al.}(2016){Soubiran}, {Le Campion}, {Brouillet}, \&
  {Chemin}}]{Soubiran:2016}
{Soubiran}, C., {Le Campion}, J.-F., {Brouillet}, N., \& {Chemin}, L. 2016,
  \aap, 591, A118

\bibitem[{{Stassun} {et~al.}(2017{\natexlab{a}}){Stassun}, {Collins}, \&
  {Gaudi}}]{StassunGaiaPlanets:2017}
{Stassun}, K.~G., {Collins}, K.~A., \& {Gaudi}, B.~S. 2017{\natexlab{a}}, \aj,
  153, 136

\bibitem[{{Stassun} {et~al.}(2014{\natexlab{a}}){Stassun}, {Feiden}, \&
  {Torres}}]{Stassun:2014}
{Stassun}, K.~G., {Feiden}, G.~A., \& {Torres}, G. 2014{\natexlab{a}}, \nar,
  60, 1

\bibitem[{{Stassun} {et~al.}(2012){Stassun}, {Kratter}, {Scholz}, \&
  {Dupuy}}]{Stassun:2012}
{Stassun}, K.~G., {Kratter}, K.~M., {Scholz}, A., \& {Dupuy}, T.~J. 2012, \apj,
  756, 47

\bibitem[{{Stassun} {et~al.}(2014{\natexlab{b}}){Stassun}, {Pepper}, {Oelkers},
  {Paegert}, {De Lee}, \& {Sanchis-Ojeda}}]{StassunTESS:2014}
{Stassun}, K.~G., {Pepper}, J.~A., {Oelkers}, R., {et~al.} 2014{\natexlab{b}},
  ArXiv e-prints, arXiv:1410.6379

\bibitem[{{Stassun} \& {Torres}(2016{\natexlab{a}})}]{StassunGaiaEB:2016}
{Stassun}, K.~G., \& {Torres}, G. 2016{\natexlab{a}}, \aj, 152, 180

\bibitem[{{Stassun} \& {Torres}(2016{\natexlab{b}})}]{StassunGaiaError:2016}
---. 2016{\natexlab{b}}, \apjl, 831, L6

\bibitem[{{Stassun} {et~al.}(2017{\natexlab{b}}){Stassun}, {Oelkers}, {Pepper},
  {Paegert}, {De Lee}, {Torres}, {Latham}, {Muirhead}, {Dressing},
  {Rojas-Ayala}, {Mann}, {Fleming}, {Levine}, {Silvotti}, {Plavchan}, \& {the
  TESS Target Selection Working Group}}]{StassunTIC:2017}
{Stassun}, K.~G., {Oelkers}, R.~J., {Pepper}, J., {et~al.} 2017{\natexlab{b}},
  ArXiv e-prints, arXiv:1706.00495

\bibitem[{{Stevens} {et~al.}(2017){Stevens}, {Stassun}, \&
  {Gaudi}}]{Stevens:2017}
{Stevens}, D., {Stassun}, K., \& {Gaudi}, B. 2017, \apj, in press

\bibitem[{{Torres} {et~al.}(2010){Torres}, {Andersen}, \&
  {Gim{\'e}nez}}]{Torres:2010}
{Torres}, G., {Andersen}, J., \& {Gim{\'e}nez}, A. 2010, \aapr, 18, 67

\end{thebibliography}

\end{document}